\documentclass{iopart}             
\usepackage[T1]{fontenc}
\usepackage{iopams}
\usepackage{graphicx}
\usepackage{url,graphics,epsfig}
\usepackage{amssymb}
\usepackage{cite}

\begin{document}

\jl{2}

\def\etal{{\it et al~}}
\def\newblock{\hskip .11em plus .33em minus .07em}
 
\setlength{\arraycolsep}{2.5pt}             

\title{Quadrupole association and dissociation of hydrogen 
in the early universe}

\author{Robert C Forrey\footnote[1]{E-mail:rcf6@psu.edu}}

\address{Department of Physics, Pennsylvania State University, 
Berks Campus, Reading, PA 19610-6009, USA}

\begin{abstract}
Radiative association and photodissociation rates
are calculated for quadrupole transitions of H$_2$.
A complete set of bound and unbound states are included in
a self-consistent master equation to obtain steady-state
concentrations for a dilute system of hydrogen atoms and
molecules. Phenomenological rate constants computed from 
the steady-state concentrations satisfy detailed balance 
for any combination of matter and radiation temperature.
Simple formulas are derived for expressing the steady-state 
distributions in terms of equilibrium distributions. 
The rate constant for radiative association is found
to be generally small for all temperature combinations.
The photodissociation rate constant for quadrupole 
transitions is found to dominate the rate constants 
for other H$_2$ photodestruction mechanisms 
for $T_R\le$ 3000 K. Implications for the formation
and destruction of H$_2$ in the early universe are discussed.

\end{abstract}


\newpage

\section{Introduction}

It is widely accepted that hydrogen molecules catalyzed the
formation of the first stars in the universe 
\cite{Saslaw1967,Peebles1968,Palla1983,Lepp2002,Kreckel2010}. 
As primordial condensations formed, the gas density increased 
and subsequent collisions produced internally excited states
which quenched via the emission of electric quadrupole 
radiation. Although inefficient, the removal of the
quadrupole radiation from the condensation allowed
gravitational collapse to proceed toward the eventual
first star. 
The probability of a quadrupole transition scales as the 
fifth power of the difference in internal energy levels \cite{kate,ionel}.
Therefore, unbound and quasibound states which provide
large energy differences might play a role in the
early stages of cooling.
Nevertheless, the possibility that hydrogen atoms
could form molecules via the emission of quadrupole
radiation has so far been neglected.
Considering that the main processes leading to the
formation of H$_2$ are known to be quite slow, 
it is worth investigating the quadrupole association
mechanism to see whether it makes a significant contribution
to the formation rate.
Such a study is particularly compelling in light of recent
developments \cite{Forrey2015} concerning the role of quasibound 
states in the recombination process. It was 
shown for conditions of local thermodynamic equilibrium (LTE)
that long-lived orbiting resonances can enhance
the radiative asssociation (RA) rate constant by several
orders of magnitude \cite{sio} compared to calculations where 
they are neglected. Hydrogen interactions allow 
orbiting resonances whose lifetimes are extremely long
(some are comparable to the age of the universe \cite{schwenke}).
Such resonances can provide feedback which may further
enhance the formation rates when the gas is not in LTE 
as is the case in the early universe after the matter
and radiation temperatures have decoupled. Therefore,
in the present work, we provide a complete account of all 
resonant and non-resonant contributions for the RA process
\begin{equation}
H+H\rightarrow H_2+\nu
\end{equation}
where $\nu$ represents the photon released in a
quadrupole transition.

The inverse process of photodissociation via 
quadrupole transitions is also considered. 
Photodissociation of H$_2$ is believed to be important 
in the early universe. Direct collapse to a supermassive 
black hole may occur if there are no hydrogen molecules 
to promote cooling and fragmentation \cite{Bromm2003}.
A strong radiation field, such as may occur after 
ignition of the first stars, destroys the formation
of H$^-$ and H$_2^+$ which are precursors for the
formation of H$_2$.  Direct and indirect photodissociation
via dipole transitions to the Lyman and Werner states 
deplete H$_2$ at high energies. 
The critical value of the Lyman and Werner intensity 
$J_{21}^{crit}$ needed for supermassive star formation
\cite{Sugimura2014,Sugimura2016}
is determined by the relative strength of the rate constants 
for photodestruction of H$^-$, H$_2^+$, and H$_2$.
Because quadrupole radiation allows transitions within 
the same ground electronic state, the mechanism would 
apply at both low and high radiation temperatures and 
perhaps contribute to the depletion of H$_2$.

In order to perform the study, we use the self-consistent
quantum kinetic theory developed previously \cite{Forrey2015}
which computes the rate constants from the steady-state solution 
of a Sturmian master equation. The bound and unbound energy 
eigenstates of the Sturmian representation form a complete 
basis set for both the dynamics and kinetics. Consequently,
all transitions between bound and unbound states,
which include long-lived quasibound and discretized 
non-resonant states, are accounted for. 
The summed rate coefficients are 
then used to provide phenomenological
rate constants for an idealized system
consisting of dilute hydrogen atoms and molecules
which undergo negligible inelastic collisions.
The RA and photodissociation rate constants computed 
in this way are guaranteed to satisfy detailed balance 
for any combination of matter and radiation temperature.
Therefore, these rate constants are suitable for use
in an expanded network of chemical processes. In the
present work, however, we consider only the quadrupole
mechanism for association and dissociation of hydrogen
and draw conclusions about its importance in the early
universe.


\section{Theory}

We begin by considering the rate equation
\begin{equation}
\frac{d}{dt}[H_2]=M_r[H]^2-M_d[H_2]
\label{rate}
\end{equation}
for a fixed number density of hydrogen nuclei
\begin{equation}
n_H=[H]+2[H_2]
\end{equation}
where square brackets indicate the concentration
of the enclosed species, and $M_r$ and $M_d$ are
the respective rate constants for recombination
and dissociation.  The solution to the rate
equation may be written
\begin{equation}
[H_2]=\frac{\left(\frac{n_H}{2}\right)^2(1-e^{-\lambda t})}
{\alpha+\beta+(\alpha-\beta)e^{-\lambda t}}
\label{H2}
\end{equation}
\begin{equation}
\alpha=\frac{\lambda}{8M_r}
\end{equation}
\begin{equation}
\beta=\frac{4M_rn_H+M_d}{8M_r}
\end{equation}
\begin{equation}
\lambda=(8M_rM_dn_H+M_d^2)^{1/2}\ .
\end{equation}
The rate constants in equation (\ref{rate}) are calculated
using the quantum kinetic theory described previously \cite{Forrey2015}.
This method requires that the single rate equation (\ref{rate}) be
consistent with a Sturmian master equation which contains 
all possible transitions, including those involving the discretized
continuum. In general, the state-to-state 
transition probability is given by
\begin{equation}
M_{i\rightarrow j}=\left\{\begin{array}{ll}
A_{i\rightarrow j}+B_{i\rightarrow j}\overline{J}_{i\rightarrow j}
+C_{i\rightarrow j} & \ \ E_i>E_j \nonumber\\
\nonumber\\
B_{i\rightarrow j}\overline{J}_{i\rightarrow j}+C_{i\rightarrow j} 
& \ \ E_j>E_i \end{array}\right. 
\label{mij}
\end{equation}
where $A$ and $B$ are Einstein coefficients, 
$C$ is a collisional rate coefficent, and $\overline{J}$
is the average radiation field at the frequency
corresponding to a molecular transition
between states with energies $E_i$ and $E_j$.
The master equation may be divided into 
two sets of equations corresponding to bound ($b$)
and unbound ($u$) states
\begin{eqnarray}
\frac{d}{dt}[H_2(b_i)] &=& \sum_j
\left(M_{u_j\rightarrow b_i}[H_2(u_j)]
-M_{b_i\rightarrow u_j}[H_2(b_i)]\right)\nonumber\\
&+& \sum_{j\neq i}\left(M_{b_j\rightarrow b_i}[H_2(b_j)]
-M_{b_i\rightarrow b_j}[H_2(b_i)]
\right)
\label{rateb}
\end{eqnarray}
\begin{eqnarray}
\frac{d}{dt}[H_2(u_i)] &=& \sum_j
\left(M_{b_j\rightarrow u_i}[H_2(b_j)]
-M_{u_i\rightarrow b_j}[H_2(u_i)]\right)\nonumber\\
&+& \sum_{j\neq i}\left(M_{u_j\rightarrow u_i}[H_2(u_j)]
-M_{u_i\rightarrow u_j}[H_2(u_i)]\right)\nonumber\\
&+&k_{f\rightarrow u_i}^{elastic}[H]^2-\tau^{-1}_{u_i}[H_2(u_i)]
\label{rateu}
\end{eqnarray}
where $\tau_{u_i}$ is the lifetime of the unbound state $u_i$,
and $k_{f\rightarrow u_i}^{elastic}$ is the two-body elastic scattering
rate constant for a discretized free ($f$) state, which is related to 
$\tau_{u_i}^{-1}$ by the equilibrium constant
\begin{equation}
K_{u_i}^{eq}=\frac{g_{u_i}\exp(-E_{u_i}/k_BT)}{Q_H^2Q_T}\ .
\label{keq}
\end{equation}
$Q_H=4$ is the atomic partition function\footnote{For the temperatures
considered in the present work, the H atom is assumed to be in the ground state,
and the atomic partition function is due to electron and nuclear spin degeneracy.},
$g_{u}=(2I_u+1)(2j_u+1)$ is the degeneracy of the unbound state
($I_u=0$ for para-H$_2$ and $I_u=1$ for ortho-H$_2$),
$Q_T$ is the translational partition function for 
temperature $T$, and $k_B$ is Boltzmann's constant.

\newpage

The steady-state solution to the master equation yields 
the rate constants
\begin{equation}
M_r=\frac{\sum_{ij}g_{u_i}e^{-E_{u_i}/k_BT}
(1+\delta_{u_i})
M_{u_i\rightarrow b_j}^{\mbox{\tiny{OUT}}}}
{Q_H^2Q_{T}}
\label{mr}
\end{equation}
and
\begin{equation}
M_d=\frac{\sum_{ij}g_{b_i}e^{-E_{b_i}/k_BT}
(1+\delta_{b_i})
M_{b_i\rightarrow u_j}^{\mbox{\tiny{OUT}}}}{Q_{H_2}}
\label{md}
\end{equation}
where
\begin{equation}
Q_{H_2}=\sum_i (1+\delta_{b_i})g_{b_i}\exp(-E_{b_i}/k_BT)
\end{equation}
is the molecular partition function, 
$\delta_b$ and $\delta_u$ are non-LTE concentration
defects which may be obtained from the steady-state
concentrations
\begin{equation} 
1+\delta_{b_i}=
\frac{\left(\frac{Q_{H_2}[H_2(b_i)]}{[H_2]}\right)_{\mbox{\tiny{SS}}}}
{\left(\frac{Q_{H_2}[H_2(b_i)]}{[H_2]}\right)_{\mbox{\tiny{LTE}}}}
=\sum_{j,k,l}\tilde{A}_{b_ib_j}^{-1}
M_{b_j\rightarrow u_k}^{\mbox{\tiny{IN}}}\tilde{B}_{u_ku_l}^{-1}
\label{deltab}
\end{equation}
\begin{equation}
1+\delta_{u_i}=
\frac{\left(\frac{[H_2(u_i)]}{[H]^2}\right)_{\mbox{\tiny{SS}}}}
{\left(\frac{[H_2(u_i)]}{[H]^2}\right)_{\mbox{\tiny{LTE}}}}
=\sum_j\tilde{B}_{u_iu_j}^{-1}
\label{deltau}
\end{equation}
and the matrices $\tilde{A}$ and $\tilde{B}$
and the rate coefficients
$M_{i\rightarrow j}^{\mbox{\tiny{IN}}}$ and
$M_{i\rightarrow j}^{\mbox{\tiny{OUT}}}$
are defined below.
Equations (\ref{deltab}) and (\ref{deltau}) give the identity       
\begin{eqnarray}
1=\frac{\sum_{ij}g_{u_i}e^{-E_{u_i}/k_BT}
(1+\delta_{u_i})
M_{u_i\rightarrow b_j}^{\mbox{\tiny{OUT}}}}
{\sum_{ij}g_{b_i}e^{-E_{b_i}/k_BT}
(1+\delta_{b_i})
M_{b_i\rightarrow u_j}^{\mbox{\tiny{OUT}}}}
\end{eqnarray}
so that the ratio
\begin{equation}
\frac{M_r}{M_d}=\frac{[H_2]}{[H]^2}
=\frac{Q_{H_2}}{Q_H^2Q_T}
\label{saha}
\end{equation}
reduces to the statistical Saha relation when 
all $\delta_{b}=\delta_{u}=0$. 
The matrices $\tilde{A}$ and $\tilde{B}$ which 
need to be inverted are obtained from equations
(\ref{rateb}) and (\ref{rateu}) and are given by
\begin{eqnarray}
\tilde{A}_{b_ib_j}&=&\delta_{ij}\left(\sum_{k\neq i}
M_{b_i\rightarrow b_k}^{\mbox{\tiny{OUT}}}+\sum_k
M_{b_i\rightarrow u_k}^{\mbox{\tiny{OUT}}}\right)
-(1-\delta_{ij})M_{b_i\rightarrow b_j}^{\mbox{\tiny{IN}}}
\label{amat}
\end{eqnarray}
\begin{eqnarray}
\tilde{B}_{u_iu_j}&=&\delta_{ij}\left\{1+\tau_{u_i}\left(
\sum_{k\neq i}M_{u_i\rightarrow u_k}^{\mbox{\tiny{OUT}}}
+\sum_kM_{u_i\rightarrow b_k}^{\mbox{\tiny{OUT}}}
\right)\right\}\nonumber\\
&-& \tau_{u_i}\left\{(1-\delta_{ij})
M_{u_i\rightarrow u_j}^{\mbox{\tiny{IN}}}
+\sum_{k,l}M_{u_i\rightarrow b_k}^{\mbox{\tiny{IN}}}
\tilde{A}_{b_kb_l}^{-1}M_{b_l\rightarrow u_j}^{\mbox{\tiny{IN}}}\right\}\ .
\label{bmat}
\end{eqnarray}
In the present work, we exclude inelastic collisions so that
$C_{i\rightarrow j}\equiv 0$ in equation (\ref{mij}).
Assuming a complete redistribution of frequencies \cite{hummer},
the average radiation field may be written in terms of the
Planck black-body radiation function, $P_{i\rightarrow j}$, 
and a line source function, $S_{i\rightarrow j}$, 
as follows \cite{alex}
\begin{equation}
\overline{J}_{i\rightarrow j}(T_M,T_R)=
(1-\beta_{ij})S_{i\rightarrow j}(T_M)
+\beta_{ij}P_{i\rightarrow j}(T_R)
\end{equation}
where $\beta_{ij}$ is the escape probability for a photon
associated with the $i\rightarrow j$ transition,
and the dependence of the
matrix elements on matter temperature $T_M$ and radiation temperature $T_R$
is now shown. 
The line source function may be written
\begin{equation}
S_{i\rightarrow j}(T_M)=F(\nu_{ij})
\left(\frac{g_in_j}{g_jn_i}-1\right)^{-1}
\label{source}
\end{equation}
where $n_i$ and $n_j$ are level densities,
$\nu_{ij}$ is the frequency of the line photon,
and $F(\nu)$ is the usual Planck function coefficient.
Because the source function (\ref{source})
is derived for conditions of statistical equilibrium,
it does not contribute to the matrix elements in the 
steady-state equations (\ref{amat}) and (\ref{bmat}).
These matrices are then obtained from the Planck function
and the Maxwell-Boltzmann factor arising from elastic collisions
and may be written as
\begin{equation}
M_{i\rightarrow j}^{\mbox{\tiny{IN}}}(T_M,T_R)=\frac{g_j}{g_i}
e^{-(E_j-E_i)/k_BT_M}M_{j\rightarrow i}^{\mbox{\tiny{OUT}}}(T_R)
\end{equation}
and
\begin{equation}
M_{i\rightarrow j}^{\mbox{\tiny{OUT}}}(T_R)=\beta_{ij}R_{i\rightarrow j}(T_R)
\label{mout}
\end{equation}
with
\begin{equation}
R_{i\rightarrow j}(T_R)=\frac{A_{i\rightarrow j}}
{1-e^{-(E_i-E_j)/k_BT_R}}\ .
\end{equation}
There is no $T_M$-dependence in the outgoing rate coefficient (\ref{mout})
due to the assumption that inelastic collisions are negligible.
In order to determine $\delta_{b_i}$ and $\delta_{u_i}$, 
it is necessary to compute $\tilde{A}$ and $\tilde{B}$.
In practical computations, equations (\ref{amat}) and (\ref{bmat})
tend to suffer round-off errors, particularly at low temperatures
where the probability for upward transitions can be very small.
For such conditions, the inversion of $\tilde{A}$ may be handled analytically  
\begin{eqnarray}
y_{b_iu_j}&=&\sum_k \tilde{A}_{b_ib_k}^{-1}
M_{b_k\rightarrow u_j}^{\mbox{\tiny{IN}}}\\ 
&\approx & \frac{e^{-E_{b_i}(1/k_BT_R-1/k_BT_M)}
\sum_k g_{b_k}e^{-E_{b_k}/k_BT_M}
M_{b_k\rightarrow u_j}^{\mbox{\tiny{IN}}}}
{\sum_{kl}g_{b_k}e^{-E_{b_k}/k_BT_R}
M_{b_k\rightarrow u_l}^{\mbox{\tiny{OUT}}}
\,\delta_{I_iI_k}}\ .
\label{ymat}
\end{eqnarray}
The $\delta$-function in the denominator arises from
the impossibility of an ortho-para transition for 
quadrupole radiation. If three-body collisions
were included in $M_{i\rightarrow j}$, equation (\ref{ymat}) would
need to be modified to include inelastic and exchange reactions. 
This problem was considered previously \cite{Forrey2015} 
where it was shown that all of the unbound states
reach a steady-state concentration that is well-approximated 
by a Boltzmann distribution whenever three-body collisions are 
important.  In the present work, the approximate formula (\ref{ymat}) 
is expected to be reliable when $M_{b\rightarrow u}^{\mbox{\tiny{OUT}}}$
is less than $M_{b\rightarrow b'}^{\mbox{\tiny{OUT}}}$. 
This condition, which is nearly always met, yields the result
\begin{equation}
1+\delta_{b_i}=C_{I_i}e^{-E_{b_i}(1/k_BT_R-1/k_BT_M)}
\label{deltab2}
\end{equation}
where 
\begin{equation}
C_{I_i}=\frac{\sum_{jk}g_{u_j}e^{-E_{u_j}/k_BT_M}
(1+\delta_{u_j})
M_{u_j\rightarrow b_k}^{\mbox{\tiny{OUT}}}\,\delta_{I_iI_j}}
{\sum_{jk}g_{u_j}e^{-E_{u_j}/k_BT_R}
M_{u_j\rightarrow b_k}^{\mbox{\tiny{OUT}}}\,\delta_{I_iI_j}}
\label{symmetry}
\end{equation}
is a symmetry constant for para-H$_2$ and ortho-H$_2$
which depends only on $T_M$ and $T_R$.
This allows equation (\ref{md}) to be written
\begin{equation}
M_d=\frac{\sum_{ij}C_{I_i}g_{b_i}e^{-E_{b_i}/k_BT_R}
M_{b_i\rightarrow u_j}^{\mbox{\tiny{OUT}}}}
{\sum_i C_{I_i}g_{b_i}e^{-E_{b_i}/k_BT_R}}\ .
\label{md2}
\end{equation}
All of the $T_M$-dependence in the photodissociation rate constant 
(\ref{md2}) is contained in the symmetry constant. If $C_I$ is
approximately the same for both nuclear symmetries, then 
it is a good approximation to
use a Boltzmann distribution of states for $T=T_R$.
The $\delta_{u_i}$ may be obtained by substituting 
equation (\ref{ymat}) into equation (\ref{bmat}) to yield
\begin{eqnarray}
\tilde{B}_{u_iu_j}&=&\delta_{ij}\left\{1+\tau_{u_i}\left(
\sum_{k\neq i}M_{u_i\rightarrow u_k}^{\mbox{\tiny{OUT}}}
+(1-\epsilon_{i})\sum_kM_{u_i\rightarrow b_k}^{\mbox{\tiny{OUT}}}
\right)\right\}\nonumber\\
&-& (1-\delta_{ij})\tau_{u_i}
\left(M_{u_j\rightarrow u_i}^{\mbox{\tiny{OUT}}}
+\epsilon_{i}\sum_{k}M_{u_j\rightarrow b_k}^{\mbox{\tiny{OUT}}}\right)
\nonumber\\
&\times &\frac{g_{u_j}}{g_{u_i}} e^{-(E_{u_j}-E_{u_i})/k_BT_M}
\label{bmat2}
\end{eqnarray}
%
where
\begin{equation}
\epsilon_i=\frac{g_{u_i}e^{-E_{u_i}/k_BT_R}
\sum_k M_{u_i\rightarrow b_k}^{\mbox{\tiny{OUT}}}}
{\sum_j g_{u_j}e^{-E_{u_j}/k_BT_R}\sum_k
M_{u_j\rightarrow b_k}^{\mbox{\tiny{OUT}}}
\,\delta_{I_iI_j}}\ .
\label{epsilon}
\end{equation}
Equations (\ref{deltau}), (\ref{bmat2}), and (\ref{epsilon})
may be used to express the system of equations for the
non-LTE defects in the form
\begin{equation}
1+\delta_{u_i}=\frac{1+\tau_{u_i}\Psi_i^{\mbox{\tiny{IN}}}}
{1+\tau_{u_i}\Psi_i^{\mbox{\tiny{OUT}}}}
\end{equation} 
where
\begin{eqnarray}
\Psi_i^{\mbox{\tiny{IN}}}&=&\sum_{j\ne i}(1+\delta_{u_j})
\frac{g_{u_j}}{g_{u_i}} e^{-(E_{u_j}-E_{u_i})/k_BT_M}\nonumber\\
&\times & \left(M_{u_j\rightarrow u_i}^{\mbox{\tiny{OUT}}}
+\epsilon_{i}\sum_{k}M_{u_j\rightarrow b_k}^{\mbox{\tiny{OUT}}}\right)
\end{eqnarray}
and
\begin{equation}
\Psi_i^{\mbox{\tiny{OUT}}}=
\sum_{k\neq i}M_{u_i\rightarrow u_k}^{\mbox{\tiny{OUT}}}
+(1-\epsilon_{i})\sum_kM_{u_i\rightarrow b_k}^{\mbox{\tiny{OUT}}}\ .
\end{equation}
In the limiting case $T_R\rightarrow 0$, equation (\ref{epsilon})
yields $\epsilon_1=1$ for the lowest energy unbound state $u_1$
and $\epsilon_i=0$ for all states with energy $E_{u_i}>E_{u_1}$.
If $\epsilon_i$ is set to zero for all $i$, then the above formulas 
reduce to ones given previously \cite{Forrey2015}
\begin{eqnarray}
\Psi_i^{\mbox{\tiny{IN}}}&=&\sum_{j>i}(1+\delta_{u_j})
\frac{g_{u_j}}{g_{u_i}} e^{-(E_{u_j}-E_{u_i})/k_BT_M}
A_{u_j\rightarrow u_i}
\end{eqnarray}
and
\begin{equation}
\Psi_i^{\mbox{\tiny{OUT}}}=
\sum_{j<i}A_{u_i\rightarrow u_j}
+\sum_j A_{u_i\rightarrow b_j}
\end{equation}
which may be solved iteratively starting with the highest energy
quasibound state and assuming that $\delta_{u_j}=0$ for all unbound states 
that are not quasibound. If $\delta_{u_j}=-1$ 
for all unbound states with $j>i$, then the formula
\begin{equation}
1+\delta_{u_i}=\frac{1}{1+\tau_{u_i}\left(
\sum_{j<i}A_{u_i\rightarrow u_j}+\sum_j A_{u_i\rightarrow b_j}\right)}
\label{approx}
\end{equation}
is obtained which may be substituted into equation (\ref{mr}) 
to yield the commonly-used Bain and Bardsley formula \cite{Bain1972}
for resonant radiative association.

\section{Results}

The bound and unbound states of H$_2$ were obtained by
solving the radial Schr\"odinger equation
\begin{eqnarray}
\left[\frac{1}{2\mu}\,\frac{d^2}{dr^2}-\frac{j(j+1)}{2\mu\,r^2}
-v(r)+E_{vj}\right]\,\chi_{vj}(r)=0\ ,
\label{diatom}
\end{eqnarray}
where $\mu$ is the reduced mass and $v(r)$ is the H$_2$
potential of Schwenke \cite{schwenke}.
The notation $b$ and $u$ is understood to mean a unit-normalized
bound or unbound energy eigenstate characterized by the pair
of quantum numbers $(v,j)$ where $v$ and $j$ are the 
vibrational and rotational quantum
numbers for the eigenstate $\chi_{vj}$.
These eigenstates were obtained by diagonalization 
in an orthonormal $L^2$ Sturmian basis set consisting
of Laguerre polynomial functions as described previously \cite{Forrey2013}.
The vibrational quantum number for an unbound eigenstate 
corresponds to the quadrature index and only 
has meaning with respect to the scale factor 
and number of basis functions.

The Einstein A-coefficient for a radiative transition
between an initial state $i$ and a final state $f$ 
is given by \cite{kate,ionel}
\begin{equation}
A_{i\rightarrow f}=1.4258\times 10^{4}
(E_i-E_f)^5 |\langle\chi_i|Q(R)|\chi_f\rangle|^2 
f(j_i,j_f)\ \mbox{s}^{-1}
\label{einstein}
\end{equation}
where $R$ is the internuclear distance, 
$Q(R)$ is the quadrupole moment, $\chi_i$ and $\chi_f$
are the energy eigenfunctions of the initial and final
states, and the branching ratio is given by
\begin{equation}
f(j_i,j_f)=\left\{\begin{array}{ll}
\frac{3(j_i+1)(j_i+2)}{2(2j_i+1)(2j_i+3)} & \ \ j_f=j_i+2 \\\nonumber\\
\frac{j_i(j_i+1)}{(2j_i-1)(2j_i+3)} & \ \ j_f=j_i \\\nonumber\\
\frac{3j_i(j_i-1)}{2(2j_i-1)(2j_i+1)} & \ \ j_f=j_i-2 \end{array}\right. 
\end{equation}
The Sturmian representation was tested 
by computing the A-coefficients 
for all bound to bound transitions using the quadrupole 
moment provided by Wolniewicz et al. \cite{ionel}
Excellent agreement was found for all of the tabulated
transitions.  

Equation (\ref{einstein}) also applies to 
transitions involving resonant and non-resonant unbound states.
Tables 1-4 provide an extension of the published tables \cite{ionel}
to include transitions between bound and quasibound states.
Only narrow quasibound states that are well-represented
by a single Sturmian eigenstate are included in the tables.
Figure 1 shows the cumulative radiative width
\begin{equation}
\Gamma_u=\sum_b A_{u\rightarrow b}
\label{cumulative}
\end{equation}
for para-H$_2$ and ortho-H$_2$ formation. 
The sharp resonances due to long-lived quasibound states
are clearly evident. In virtually all previous work on RA,
extremely narrow resonances of this kind have been neglected,
either because they are difficult to resolve using a 
grid-based method or because of an assumed breakdown
of perturbation theory \cite{Bennett2003,Gustafsson2012,Antipov2013}. 
Recently, it was shown \cite{sio}
that such resonances increased the LTE rate constant 
for SiO formation by $\sim$ 100 times compared to
calculations where the resonances were neglected.
A similar observation \cite{Forrey2015} was made
for the RA rate constant for CH$^+$ where it
was also predicted that the non-LTE defects
could contribute to a further increase (or decrease)
compared to the LTE values.

To explore this possibility for H$_2$, we computed 
$M_{i\rightarrow j}^{\mbox{\tiny{IN}}}$
and $M_{i\rightarrow j}^{\mbox{\tiny{OUT}}}$ in order to
evaluate the non-LTE defects. We assume $\beta_{ij}=1$
which means there is no distortion of the background 
blackbody radiation field. This assumption is convenient
but not necessary - the formulation works as written
for an optically thick gas so long as the complete
redistribution of frequencies \cite{hummer} is assumed 
and the escape probabilities are calculated using the line
source function (\ref{source}).
Figures 2-5 show the non-LTE defects for the four
longest-lived quasibound states of hydrogen.
The tunneling widths \cite{schwenke}
for these states are
$1.1\times 10^{-21}$, 
$8.2\times 10^{-25}$,
$1.5\times 10^{-18}$, and
$6.0\times 10^{-34}$ cm$^{-1}$ for
$j$=24, 29, 31, and 32, respectively.
The curves were computed using the approximate
formula (\ref{bmat2}) and compared with the exact
equation (\ref{bmat}) for a range of temperatures
where the inversion of equation (\ref{amat}) 
was numerically stable.
Excellent agreement was found in all of the comparisons.
As expected, the plots show $\delta_u=0$ in the LTE
limit $T_R=T_M$.
Complete removal of long-lived quasibound states corresponds 
to $\delta_u=-1$. The top panels of Figures 2-5 show that this
condition is approached for low values of $T_R$ when $T_M>T_R$.
The depletion is greatest for the quasibound state
with the smallest tunneling width ($j=32$). For the
largest tunneling width shown ($j=31$), the depletion
is not as complete - the steady-state population is about
8\% of its LTE value (see Figure 4). For the majority of
quasibound states that are not shown, the steady-state
and LTE populations are nearly identical ($\delta_u\approx 0$)
for all temperatures. 

The middle and lower panels of Figures 2-5 show that $\delta_u>0$
for intermediate and large values of $T_R$ when $T_M>>T_R$. 
For para-H$_2$, the curves increase with $T_R$ before
reaching their largest values at $T_R\approx$ 200-300 K
where they begin to decrease. For ortho-H$_2$, the
pattern is similar, however, the curves reach their 
largest values at $T_R\approx 500$ K. The lower panels
for both para-H$_2$ and ortho-H$_2$ show the curves
uniformly decreasing for $T_R\ge 1000$ K.
Interestingly, Figures 2-5 show that $\delta_u>>0$
when $T_M<<T_R$ which shows that the steady-state
density of quasibound states is much greater than
the LTE density.  This suggests that
there may be a significant enhancement in the
formation rate constant when $T_M<T_R$. 

In order to see the effect of the steady-state populations,
we plot the LTE and non-LTE rate constants in Figure 6. 
The top panel shows the para-H$_2$ and ortho-H$_2$
contributions to the LTE rate constant. The large 
hump centered around 1 K is due to a low energy $j=4$ resonance
which has a tunneling lifetime of $8.4\times 10^{-6}$ cm$^{-1}$ \cite{schwenke}.
Ortho-H$_2$ formation begins to dominate at temperatures
above 100 K. The total LTE rate constant shows a broad maximum
of $\sim 10^{-28}$ cm$^3s^{-1}$ near 500 K. The middle and
lower panels of Figure 6 show how the non-LTE populations
affect the rate constant for various $T_R$. For
para-H$_2$, the steady-state and LTE contributions
are the same for $T_M<50$ K when $T_R$=10 K. As $T_M$
is increased beyond 50 K, the curve for $T_R$=10 K
drops below the LTE curve. Results for lower values 
of $T_R$ were found to be essentially the same as 
the $T_R$=10 K curve and are not shown.
For ortho-H$_2$, the steady-state and LTE contributions
are very nearly the same when $T_M$ and $T_R$ are both less
than 100 K. This is due to the somewhat larger energies for 
the ortho-H$_2$ quasibound states compared to the $j=4$ resonance.
As $T_R$ increases, the rate constants for both nuclear symmetries 
increase due to stimulated emission. The non-LTE rate constants
may be larger or smaller than their LTE values
due the behavior of $\delta_u$ shown in Figures 2-5.

The full set of $\delta_u$ may be used to compute the
symmetry constants $C_I$ defined by equation (\ref{symmetry}).
These constants allow convenient computation of the
non-LTE concentration defects $\delta_b$ from
equation (\ref{deltab2}) and the photodissociation rate
constant from equation (\ref{md2}). Numerical results
confirmed that $C_I$ is indeed a constant for all combinations
of vibrational and rotational states of a given symmetry.
The top panel of Figure 7 shows $C_I$ as a function of $T_M$
for $T_R=1000, 2000, 5000,$ and  10000 K using solid lines
for $I=0$ and dashed lines for $I=1$. A significant 
$T_M$-dependence may be seen for these constants, however, 
there is only a weak dependence on the symmetry $I$, 
particularly at high temperatures. 
Therefore, the $C_I$ in equation (\ref{md2}) approximately 
drops out which eliminates the $T_M$-dependence of the
dissociation rate constant $M_d$. This weak $T_M$-dependence is
again due to the assumption that inelastic collisions are negligible
which causes the general rate constant $M_d$ defined by equation
(\ref{md}) to be equal to the photodissociation rate constant
(\ref{md2}). In this limit, the steady-state distribution of
bound states at temperature $T_M$ is well-approximated by a
Boltzmann equilibrium distribution at temperature $T_R$.

While a Boltzmann distribution of molecular states at
temperature $T_R$ is a good approximation when inelastic
collisions are negligible, it should be noted that there 
is a scale change associated with the partition functions
\begin{equation}
\Lambda(T_R,T_M)\equiv\frac{Q_{H_2}^{SS}}{Q_{H_2}^{LTE}}
\end{equation}
which yields
\begin{equation}
\left(\frac{M_r}{M_d}\right)_{SS}=\Lambda(T_R,T_M)
\left(\frac{M_r}{M_d}\right)_{LTE}\ .
\label{scale}
\end{equation}
The bottom panel of Figure 7 shows the scale factor for
the quadrupole mechanism as a function of temperature. 
As expected, the figure shows $\Lambda=1$ when $T_R=T_M$.
However, there is a large variation in $\Lambda$ when the
temperatures are not the same. When $T_R<T_M$, the scale
factor is larger than unity, and equation (\ref{scale})
shows that the steady-state ratio $M_r/M_d$ is larger 
than the LTE ratio. This change of scale is important
whenever radiative transitions are more probable than 
inelastic collisions. In such cases, the usual detailed
balance assumption $\Lambda=1$ cannot be used.

Figure 8 shows the steady-state rate constants for 
quadrupole association and dissociation of hydrogen.
The $M_r$ curves displayed in the top panel are 
reminiscent of the RA rate constants for H and D
calculated by Stancil and Dalgarno \cite{Stancil1997}
which included stimulated plus spontaneous radiative
dipole transitions. In the present case, the rate
constants for H$_2$ formation are about 100 times 
smaller than for HD formation. The modification in
the rate constants due to the non-LTE concentration
defects $\delta_u$ is seen to be significant,
however, the net effect is still very small
due to the weak radiative coupling.
The bottom panel of Figure 8 shows the dissociation rate constant
$M_d$ as a function of $T_R$. There is no dependence on $T_M$ at
the level of resolution of the plot.
Also shown are rate constants for photoionization
\begin{equation}
H_2+\nu\rightarrow H_2^+ +e^-
\end{equation}
and indirect photodissociation
\begin{equation}
H_2+\nu\rightarrow H+H
\end{equation}
which occurs via dipole transitions to the Lyman and Werner systems.
Both of these rate constants are taken from the fitting formulas
given in Coppola et al. \cite{coppola2011}. The photodissociation
rate constant for quadrupole radiation dominates the other 
H$_2$ photodestruction rate constants for $T_R\le$ 3000 K.
For a more general radiation field comprised of low energy photons, 
the quadrupole transitions would likewise provide the strongest 
mechanism for photodestruction of H$_2$.
Because the first luminous sources produced diluted blackbody
fields with $T_R>10,000$ K, the quadrupole mechanism is not
expected to make a significant contribution to the suppression
of H$_2$ molecules.

\section{Discussion}

With the rate constants described above, we may assess 
the significance of quadrupole association and dissociation 
on the formation of H$_2$ in the early universe.
The rate constants shown in Figure 8 are
extremely small and would be unimportant in all but
the most extreme environments. The H$^-$ sequence
\begin{equation}
H+e^-\rightarrow H^- +\nu
\label{H-}
\end{equation}
\begin{equation}
H^-+H\rightarrow H_2 +e^-
\end{equation}
is known to be the dominant H$_2$ formation mechanism
in the early universe when the radiation field was weak 
\cite{Lepp2002,Kreckel2010}.
For redshift $z>100$, the cosmic background radiation
destroys H$^-$ through the reverse of process (\ref{H-}).
In this case, the RA processes
\begin{equation}
H^++H\rightarrow H_2^++\nu
\label{H2+}
\end{equation}
\begin{equation}
H^++He\rightarrow HeH^++\nu
\label{HeH+}
\end{equation}
followed by
\begin{equation}
HeH^++H\rightarrow H_2^++He
\end{equation}
\begin{equation}
H_2^++H\rightarrow H_2+H^+
\label{H+}
\end{equation}
are the dominant H$_2$ formation mechanisms.
These mechanisms are also destroyed by the cosmic background
radiation at high redshifts through the reverse of processes 
(\ref{H2+}) and (\ref{HeH+}). Because the binding energies of
H$_2^+$ and HeH$^+$ are smaller than the binding energy of H$_2$, 
they are more easily photodissociated than H$_2$, and
the quadrupole mechanisms may be significant. To see
whether this is indeed the case, we substitute 
$M_r$ and $M_d$ into the rate equation (\ref{rate})
and assume that no other processes contribute.
The steady-state solution is
\begin{equation}
[H_2]_{\mbox{\tiny{SS}}}=\left(\frac{n_H}{2}\right)^2
(\alpha+\beta)^{-1}\ .
\end{equation}
Figure 9 shows the steady-state fractional population of H$_2$
versus $n_H$ for several values of $T_R$. As $n_H$ gets large, 
the ratio increases towards a limiting value of 1/2 indicating
that the gas has become completely molecular. This is not a
realistic limit, however, because the time needed to reach
the steady-state for quadrupole association is extremely long.
The bottom panel of Figure 9 shows the time to reach one-half 
of the steady-state concentration 
\begin{equation}
t_{1/2}=\lambda^{-1}\ln\left(\frac{3\alpha+\beta}{\alpha+\beta}\right)\ .
\end{equation}
This time scale is larger than the age of the universe for low
temperatures and densities. At the higher temperatures needed to 
reach steady-state on a more realistic time scale, the fractional
population of H$_2$ drops substantially due to the increased
efficiency of photodissociation. The bottom panel of Figure 9 shows that
steady-state is obtainable within 10$^5$ years for $T_R\ge 2500$ K 
which corresponds to $n_H\approx 10^{3}$ cm$^{-3}$.
Using these numbers in the top panel shows that
the fractional population of H$_2$ is of order $10^{-12}$.
For comparison, the formation mechanisms (\ref{H2+})-(\ref{H+}) 
yield a fractional population of H$_2$ which is between 
$10^{-11}-10^{-15}$ when $500<z<1300$ \cite{coppola2011}.
The quadrupole mechanism cannot reach its steady-state limit
for $T_R<2500$ K, so it is necessary to use the time-dependent
solution (\ref{H2}) to estimate the population. The dashed curve
in the top panel of Figure 9 shows this solution for $z=500-1300$
using the usual redshift formulas
\begin{equation}
t=(14\times 10^9 \mbox{yr})(1+z)^{-3/2}
\end{equation}
\begin{equation}
T_R=(2.73 \mbox{K})(1+z)
\end{equation}
\begin{equation}
n_H=(10^{-6} \mbox{cm}^{-3})(1+z)^{3}\ .
\end{equation}
All calculations used $M_r=10^{-28}$ cm$^3/s$ so there
is no $T_M$-dependence in the plots. Figure 8 shows this 
should be a good approximation for the values of $T_R$ considered.
The fractional population is found to have a maximum near $z=800$ 
and a value of $10^{-13}$ at $z=1000$. The maximum represents
the optimal balance between the density and time needed to recombine
and the destructive efficiency of photodissociation.
The results for this restricted model demonstrate that the quadrupole 
mechanism is competitive with the standard formation mechanisms
(\ref{H2+})-(\ref{H+}) for these high redshifts
and perhaps even dominates for $800<z<1300$.

\section{Conclusions}

A theoretical formalism is described for H$_2$ which allows 
radiative association and photodissociation rate constants 
to be computed self-consistently for quadrupole transitions. 
The impact of extremely long-lived resonances is considered 
for LTE and non-LTE conditions that were prevalent in the 
early universe. Simple formulas are presented which may be 
used to compute the non-LTE concentration defects 
for any combination of matter and radiation temperature.
It is found that the bound state defects $\delta_b$ 
at temperature $T_M$ yield a good approximation to a
Boltzmann equilibrium distribution at temperature $T_R$
in the limiting case where inelastic collisions
and charge transfer reactions are negligible.
The unbound defects $\delta_u$ for the quasibound states
were found to be quite large for $T_M<<T_R$ and $T_M>>T_R$, 
but do not substantially increase the formation rate constant 
due to the large exponential decay in the Boltzmann factor at 
small $T_M$ and the increasing value of $Q_T$ at large $T_M$.
The photodissociation rate constant 
for quadrupole transitions is found to dominate the rate
constants for other H$_2$ photodestruction mechanisms 
for $T_R\le$ 3000 K. It is demonstrated that quadrupole 
association and dissociation of hydrogen is generally 
inefficient but may have occurred in the early universe for $z>500$.

\ack
The author acknowledges support from NSF Grant No. PHY-1503615.

\section*{References}
\bibliographystyle{iopart-num}
\bibliography{refs}

\begin{table*}
\centering
\caption{Einstein A-coefficients (s$^{-1}$) for radiative transitions
between quasibound and bound states of H$_2$. The rotational level $j$ 
of the quasibound state and vibrational level $v$ of the bound state 
are arranged in columns. The coefficients for transitions to
 rotational levels $j-2, j, j+2$ are shown for each $v$.  }
\label{tab1}
\begin{tabular}{llllllll}
\noalign{\vskip 1mm}
\hline  \hline \noalign{\vskip 1mm} 
$v$  & $j=4$ & $j=13$ & $j=15$ & $j=17$ & $j=19$ & $j=21$ & $j=23$  \\
\hline
  & 4.25(-13) & 1.15(-10) & 3.04(-10) & 6.35(-10) & 1.20(-9)  & 2.06(-9) & 3.03(-9) \\
0 & 1.12(-14) & 3.92(-13) & 1.94(-12) & 7.17(-12) & 2.45(-11) & 8.20(-11) & 2.83(-10) \\
  & 7.12(-13) & 4.74(-12) & 9.41(-12) & 1.69(-11) & 3.19(-11) & 6.63(-11) & 1.59(-10) \\
\hline
  & 7.46(-12) & 1.37(-9)  & 3.42(-9)  & 6.60(-9)  & 1.12(-8)  & 1.61(-8) & 1.64(-8) \\
1 & 2.44(-15) & 1.19(-11) & 4.79(-11) & 1.55(-10) & 4.80(-10) & 1.51(-9) & 5.13(-9) \\
  & 8.34(-12) & 6.52(-11) & 1.30(-10) & 2.35(-10) & 4.51(-10) & 9.60(-10) & 2.01(-9) \\
\hline
  & 6.62(-11) & 8.17(-9)  & 1.88(-8)  & 3.26(-8) & 4.69(-8) & 4.85(-8) & 1.55(-8) \\
2 & 7.42(-13) & 1.53(-10) & 5.37(-10) & 1.58(-9) & 4.62(-9) & 1.43(-8) & 4.33(-8) \\
  & 4.76(-11) & 4.43(-10) & 8.96(-10) & 1.64(-9) & 3.17(-9) & 5.61(-9) & 5.30(-9) \\
\hline
  & 3.94(-10) & 3.18(-8)  & 6.60(-8) & 9.77(-8) & 1.04(-7) & 4.07(-8) & 4.61(-8) \\
3 & 1.82(-11) & 1.18(-9)  & 3.77(-9) & 1.05(-8) & 3.01(-8) & 8.05(-8) & 1.42(-7) \\
  & 1.71(-10) & 1.98(-9)  & 4.07(-9) & 7.31(-9) & 1.10(-8) & 7.87(-9) & 1.46(-10) \\
\hline
  & 1.76(-9)  & 8.90(-8)  & 1.59(-7) & 1.76(-7) & 7.91(-8) & 3.41(-8) & 9.74(-7) \\
4 & 1.72(-10) & 6.42(-9)  & 1.97(-8) & 5.31(-8) & 1.26(-7) & 1.86(-7) & 8.13(-8) \\
  & 4.13(-10) & 6.39(-9)  & 1.24(-8) & 1.59(-8) & 8.15(-9) & 2.33(-11) & 8.08(-10) \\
\hline
  & 6.21(-9)  & 1.83(-7)  & 2.46(-7) & 1.22(-7) & 2.10(-8) & 8.44(-7) & 1.54(-6) \\
5 & 1.00(-9)  & 2.80(-8)  & 8.09(-8) & 1.72(-7) & 2.14(-7) & 7.43(-8) & 5.43(-11) \\
  & 6.44(-10) & 1.31(-8)  & 1.60(-8) & 4.90(-9) & 1.06(-9) & 1.43(-9) & \\
\hline
  & 1.80(-8)  & 2.49(-7)  & 1.50(-7)  & 1.15(-8) & 6.72(-7) & 1.08(-6) & 1.34(-7) \\
6 & 4.31(-9)  & 9.45(-8)  & 2.03(-7)  & 2.14(-7) & 5.88(-8) & 6.74(-13) & \\
  & 4.62(-10) & 7.63(-9)  & 5.15(-10) & 4.78(-9) & 2.21(-9) && \\
\hline
  & 4.46(-8) & 1.28(-7)  & 6.53(-9) & 4.89(-7) & 6.85(-7) & 6.71(-8) & \\
7 & 1.53(-8) & 1.78(-7)  & 1.80(-7) & 3.87(-8) & 6.69(-11) && \\
  & 6.13(11) & 2.07(-9)  & 1.17(-8) & 2.88(-9) &&& \\
\hline
  & 9.01(-8) & 4.65(-9)  & 3.17(-7) & 3.87(-7)  & 3.00(-8) && \\
8 & 4.41(-8) & 1.05(-7)  & 1.93(-8) & 1.88(-10) &&& \\
  & 8.28(-9) & 1.65(-8)  & 2.99(-9) &&&& \\
\hline
  & 1.15(-7) & 1.56(-7)  & 1.83(-7) & 1.14(-8) &&& \\
9 & 7.86(-8) & 5.30(-9)  & 2.64(-10) &&&& \\
  & 4.40(-8) & 1.97(-8)  &&&&& \\
\hline
   & 5.38(-8) & 5.88(-8)  & 3.36(-9) &&&& \\
10 & 5.42(-8) & 1.98(-10) &&&&& \\
   & 5.49(-8) &&&&&& \\
\hline
   & 1.70(-9) & 5.78(-10) &&&&& \\
11 & 6.54(-9) &&&&&& \\
   & 1.26(-8) &&&&&& \\
\hline
   & 1.28(-9)  &&&&&& \\
12 & 2.61(-11) &&&&&& \\
   & 1.69(-10) &&&&&& \\
\hline
   & 1.07(-10) &&&&&& \\
13 & 9.82(-12) &&&&&& \\
   & 1.07(-14) &&&&&& \\
\hline
   & 6.71(-15) &&&&&& \\
14 &&&&&&& \\
   &&&&&&& \\
\noalign{\vskip 1mm}
\hline\hline
\end{tabular}
\end{table*}

\newpage

\begin{table*}
\centering
\caption{Same as Table 1 but for $j=24-29$.}
\label{tab2}
\begin{tabular}{llllllll}
\noalign{\vskip 1mm}
\hline  \hline \noalign{\vskip 1mm}
$v$  & $j=24$ & $j=25$ & $j=26$ & $j=27$ & $j=28$ & $j=29$ & $j=29$ \\
\hline
  & 4.46(-9)  & 3.06(-9)  & 1.79(-9)  & 3.55(-10) & 8.65(-9) &  2.24(-7) & 1.95(-8) \\
0 & 1.01(-9)  & 1.08(-9)  & 4.27(-9)  & 4.44(-9) & 1.57(-8) &  3.14(-8) & 1.47(-8)  \\
  & 4.36(-10) & 4.02(-10) & 8.13(-10) & 6.45(-10) & 2.92(-10) & 2.70(-11) & 1.46(-10) \\
\hline
  & 1.14(-8) & 3.74(-9) & 1.45(-8)  & 4.06(-8) & 5.79(-7) & 2.95(-6) & 7.28(-7) \\
1 & 1.76(-8) & 1.79(-8) & 5.14(-8)  & 4.67(-8) & 5.95(-8) & 5.15(-9) & 4.32(-8) \\
  & 3.23(-9) & 2.47(-9) & 7.48(-10) & 3.37(-10) & 9.79(-11) && \\
\hline
  & 1.25(-8)  & 5.00(-8)  & 7.52(-7)  & 9.51(-7) & 3.05(-6) & 1.87(-6) & 2.89(-6) \\
2 & 1.04(-7)  & 9.17(-8)  & 9.21(-8)  & 6.44(-8) & 2.46(-9) && \\
  & 9.83(-10) & 3.67(-10) & 3.05(-10) &&&  \\
\hline
  & 8.02(-7) & 1.02(-6) & 2.75(-6) & 2.54(-6) & 9.87(-7) &  2.58(-10) & 7.48(-7) \\
3 & 1.16(-7) & 7.79(-8) & 1.92(-9) & 5.26(-10) &&&  \\
  & 7.57(-10) & 3.97(-10) &&&&& \\
\hline
  & 2.27(-6) & 2.05(-6) & 6.01(-7) & 4.40(-7) &&&  \\
4 & 1.24(-9) & 2.41(-10) &&&&& \\
  &&&&&&& \\
\hline
  & 3.53(-7)  & 2.49(-7) &&&&& \\
5 &&&&&&& \\
  &&&&&&& \\
\hline
  & 1.65(-11)  &&&&&& \\
6 &&&&&&& \\
  &&&&&&& \\
\noalign{\vskip 1mm}
\hline\hline 
\end{tabular}
\end{table*}

\begin{table*}
\centering
\caption{Same as Table 1 but for $j=30-33$.}
\label{tab3}
\begin{tabular}{llllllll}
\noalign{\vskip 1mm}
\hline  \hline \noalign{\vskip 1mm}
$v$ & $j=30$ & $j=31$ & $j=31$ & $j=32$ & $j=32$ & $j=33$ & $j=33$ \\
\hline
  & 2.93(-7) & 2.10(-6) & 3.69(-7) & 4.73(-6) & 2.15(-6) & 4.34(-6) & 2.14(-6) \\        
0 & 2.52(-8) & 3.60(-9) & 1.90(-8)  \\                                          
  & \\
\hline
  & 2.97(-6) & 2.87(-6) & 2.88(-6) & 3.60(-10) & 2.49(-6) \\             
1 & 2.59(-9)  \\                                          
  & \\
\hline
  & 1.58(-6) \\
2 & \\
  & \\
\noalign{\vskip 1mm}
\hline\hline
\end{tabular}
\end{table*}

\begin{table*} 
\centering 
\caption{Einstein A-coefficients (s$^{-1}$) for radiative transitions
between quasibound states of H$_2$. For rotational levels with more
than one quasibound state, the entries are listed according to 
$j_{u_f}=j_{u_i}-2$ starting with the lowest $E_{u_i}$ and $E_{u_f}$, 
and ending with the $j_{u_f}=j_{u_i}$ transition. }
\label{tab4}
\begin{tabular}{llllllllll}
\noalign{\vskip 1mm}
\hline  \hline \noalign{\vskip 1mm}
$j_{u_i}$ & $A_{u_i\rightarrow u_f}$ & $j_{u_i}$ & $A_{u_i\rightarrow u_f}$ &
$j_{u_i}$ & $A_{u_i\rightarrow u_f}$ & $j_{u_i}$ & $A_{u_i\rightarrow u_f}$ &
$j_{u_i}$ & $A_{u_i\rightarrow u_f}$ \\
\hline
15 & 1.29(-20) & 26 & 7.43(-11) & 30 & 5.08(-10) & 32 & 1.88(-11) & 33 & 1.47(-9)  \\
17 & 4.50(-16) & 27 & 1.83(-10) & 31 & 4.96(-10) & 32 & 8.74(-10) & 33 & 8.56(-10) \\
19 & 8.10(-14) & 28 & 2.25(-10) & 31 & 5.17(-11) & 32 & 1.94(-9)  & 34 & 3.88(-6)  \\
21 & 1.60(-12) & 29 & 1.55(-10) & 31 & 1.24(-6)  & 33 & 6.14(-10) & 34 & 1.01(-9)  \\
23 & 1.24(-11) & 29 & 4.56(-10) & 31 & 9.19(-10) & 33 & 1.11(-11) & 35 & 3.34(-6)  \\
25 & 5.66(-11) & 29 & 8.26(-10) & 31 & 1.01(-9)  & 33 & 2.04(-6)  & 35 & 1.64(-9)  \\
\noalign{\vskip 1mm}
\hline\hline
\end{tabular}
\end{table*} 

\newpage

\begin{figure}
\centerline{\epsfxsize=4in\epsfbox{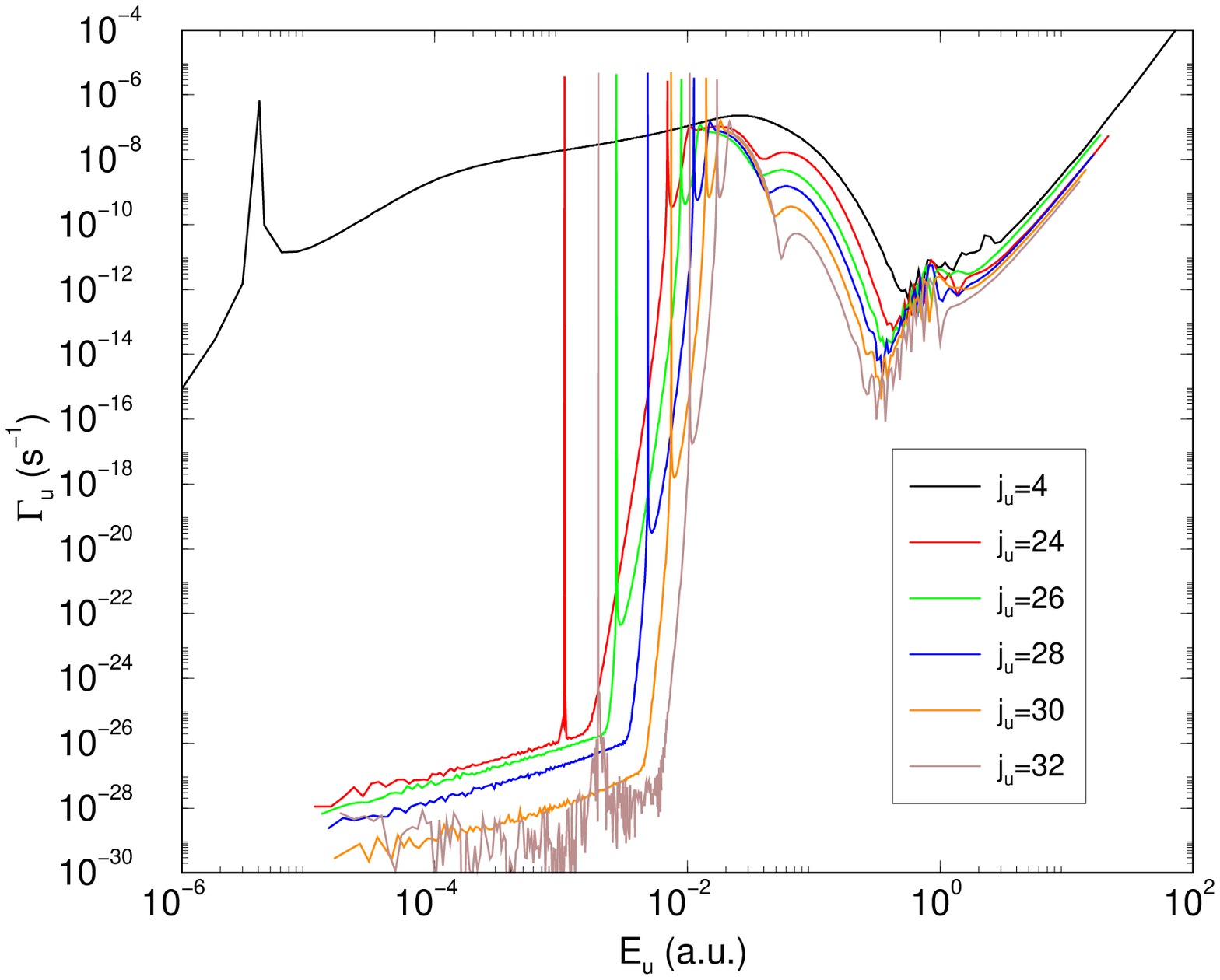}}
\centerline{\epsfxsize=4in\epsfbox{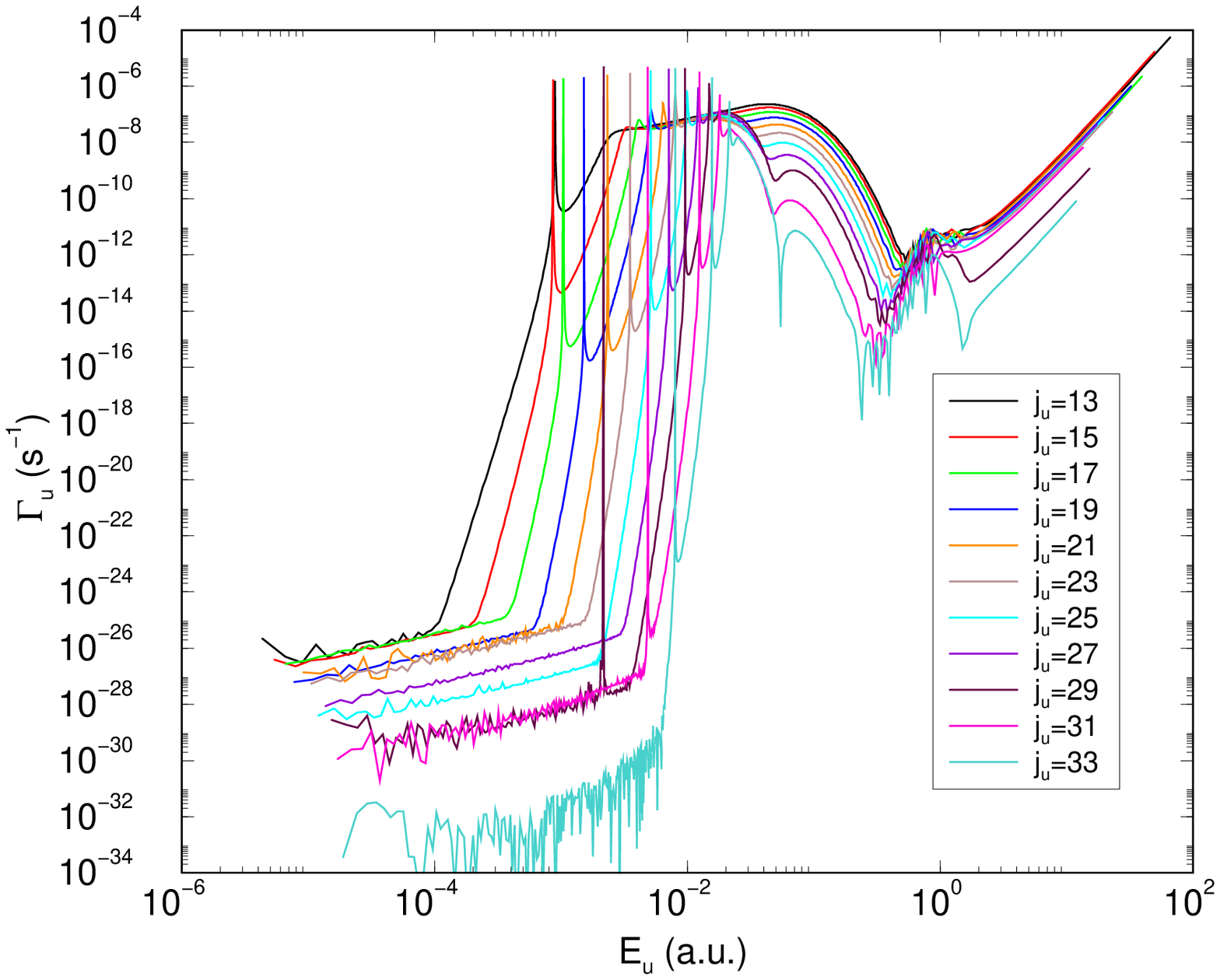}}
\caption{Radiative association width (\ref{cumulative})
for the formation of para-H$_2$ (top) and ortho-H$_2$ (bottom).
Partial waves that support long-lived quasibound states are shown.
The resonance peaks correspond to summed columns in Tables 1-3.
}
\label{fig1}
\end{figure}

\newpage

\begin{figure}
\centerline{\epsfxsize=4in\epsfysize=2.6in\epsfbox{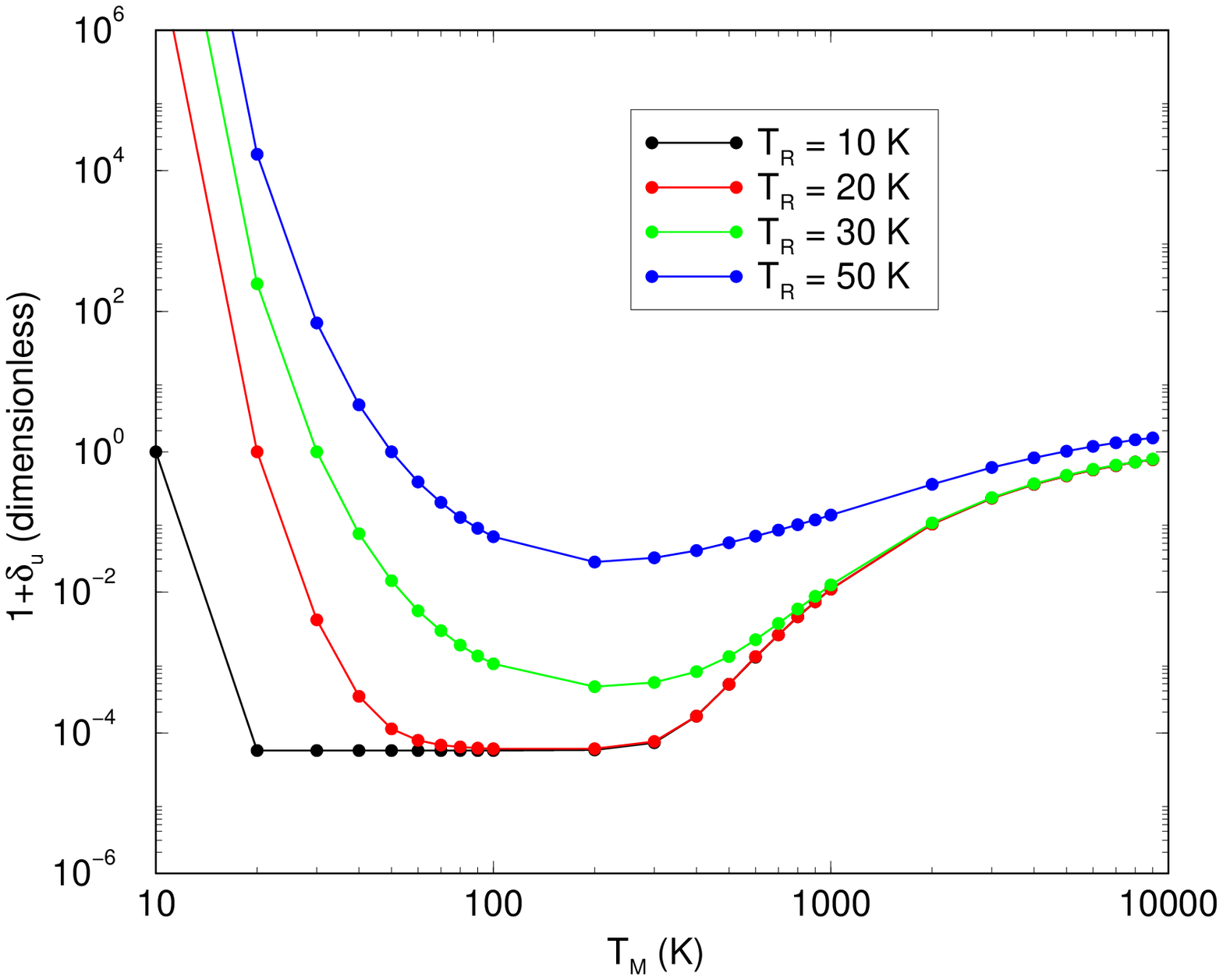}}
\centerline{\epsfxsize=4in\epsfysize=2.6in\epsfbox{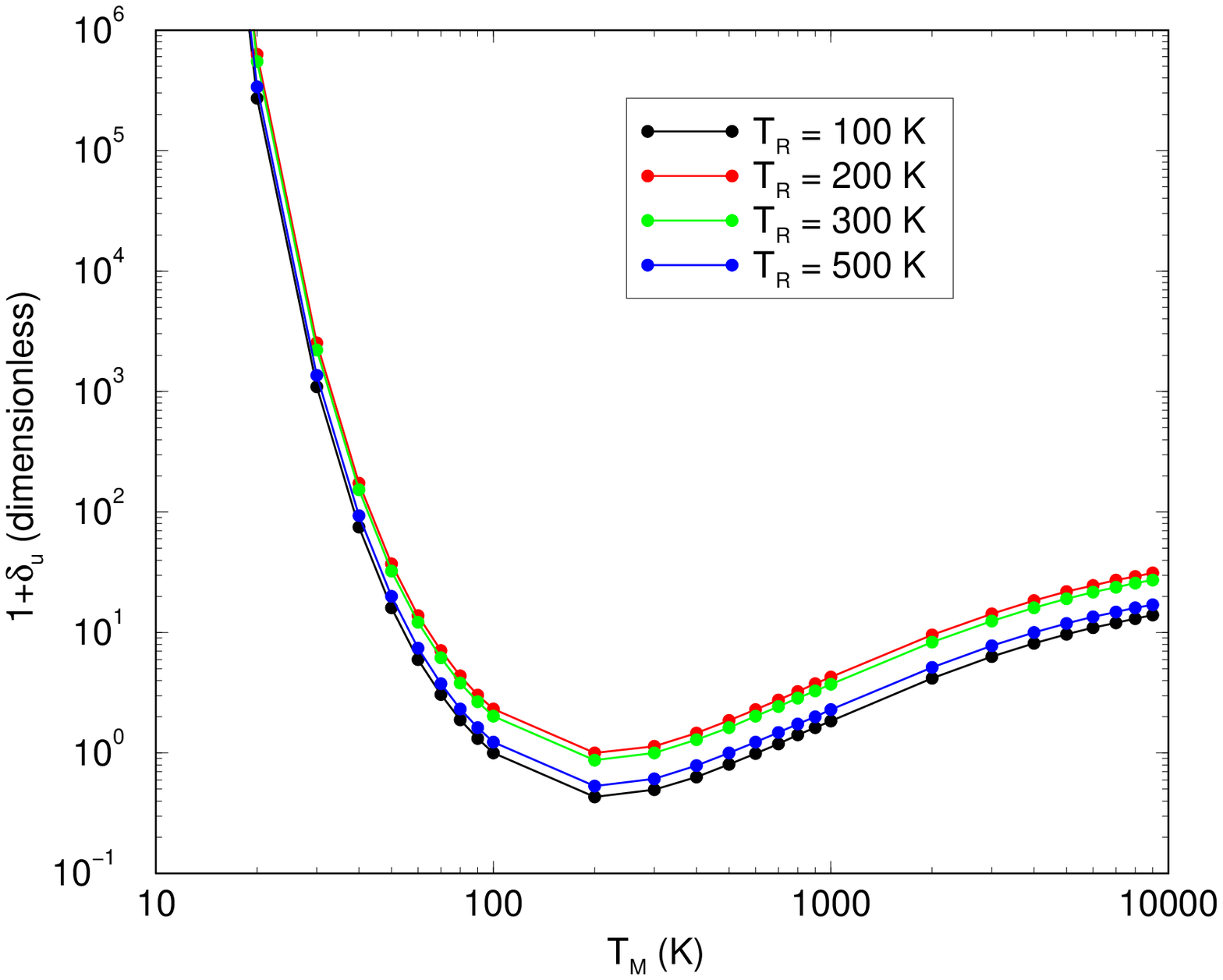}}
\centerline{\epsfxsize=4in\epsfysize=2.6in\epsfbox{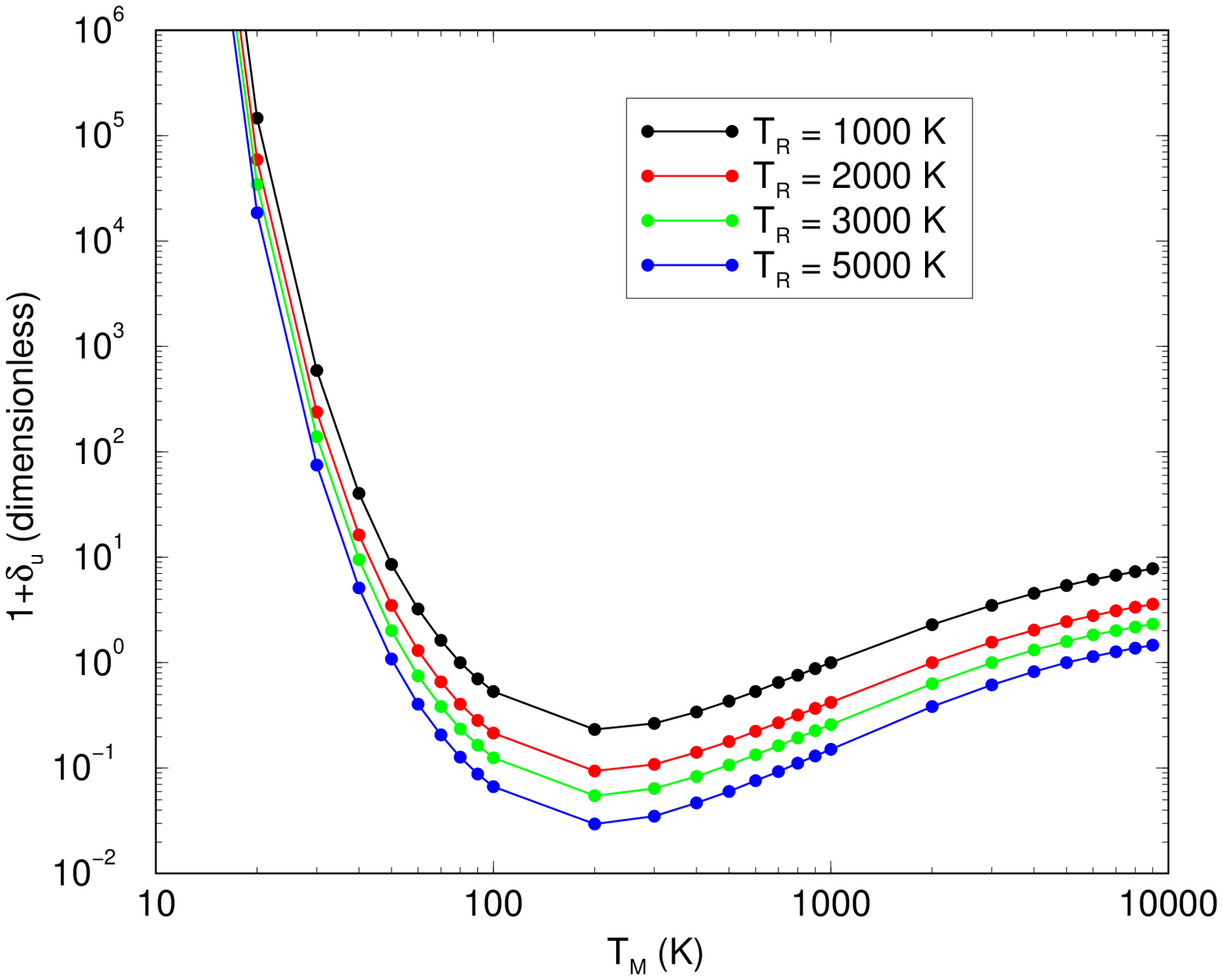}}
\caption{Non-LTE concentration factor for the $j$=24 quasibound state 
of $H_2$. Low values of $T_R$ tend to deplete  
the resonant state for $T_M>T_R$. Intermediate values of $T_R$
tend to increase the quasibound concentration beyond the LTE value,
especially at low $T_M$. Large values of $T_R$ tend to deplete the
quasibound state at intermediate $T_M$ while enhancing 
the concentration at low and high $T_M$.}
\label{fig2}
\end{figure}

\newpage

\begin{figure}
\centerline{\epsfxsize=4in\epsfysize=2.6in\epsfbox{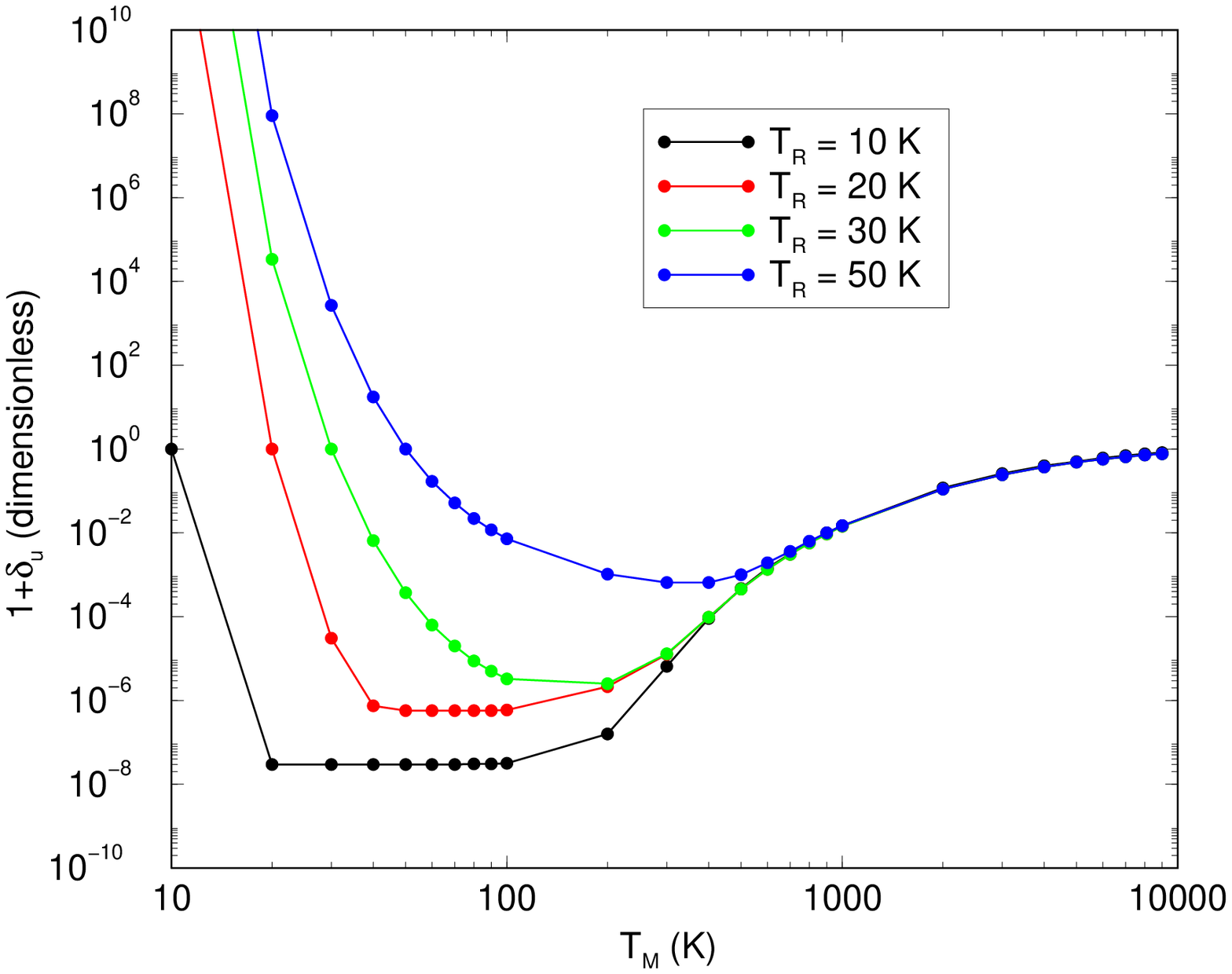}}
\centerline{\epsfxsize=4in\epsfysize=2.6in\epsfbox{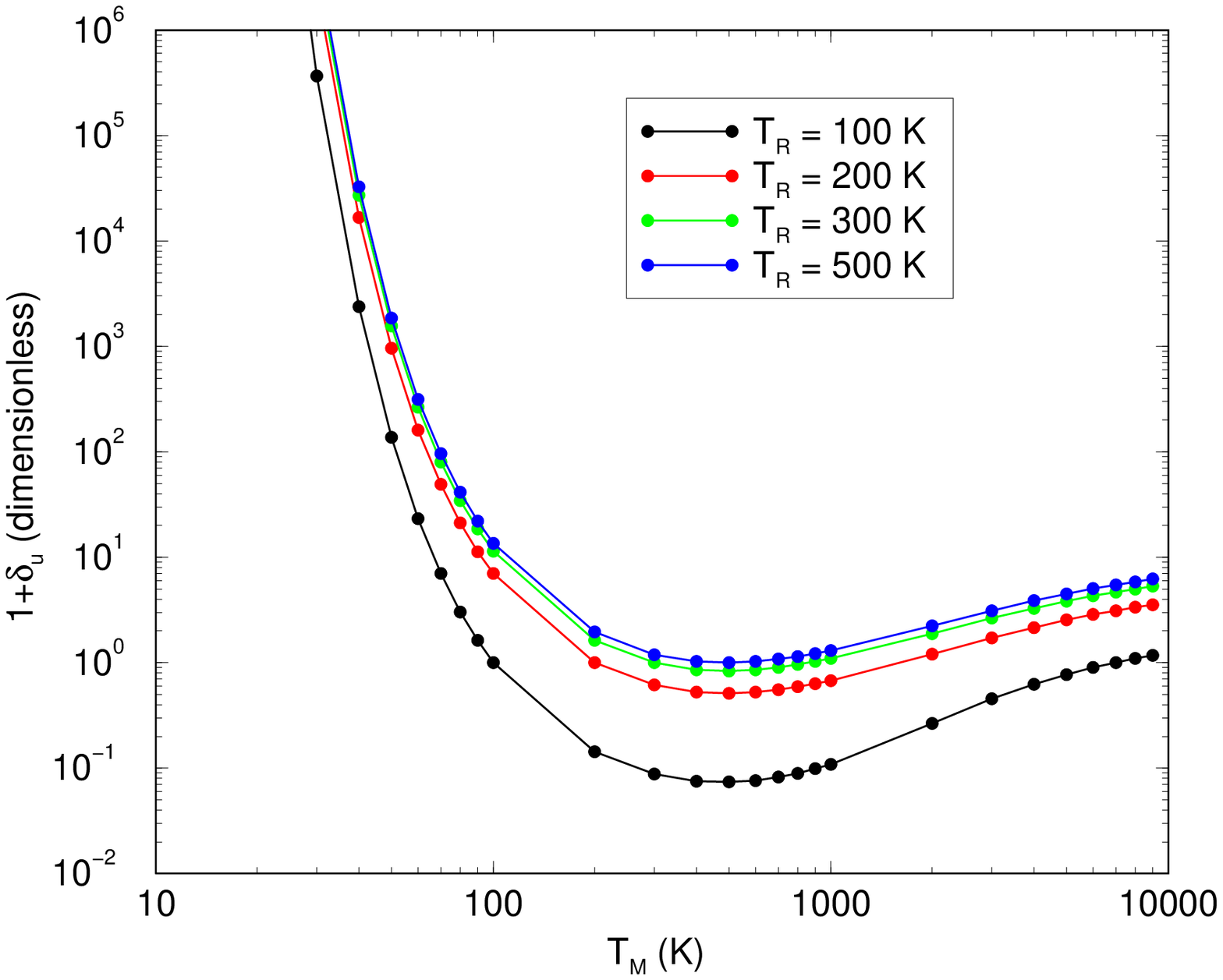}}
\centerline{\epsfxsize=4in\epsfysize=2.6in\epsfbox{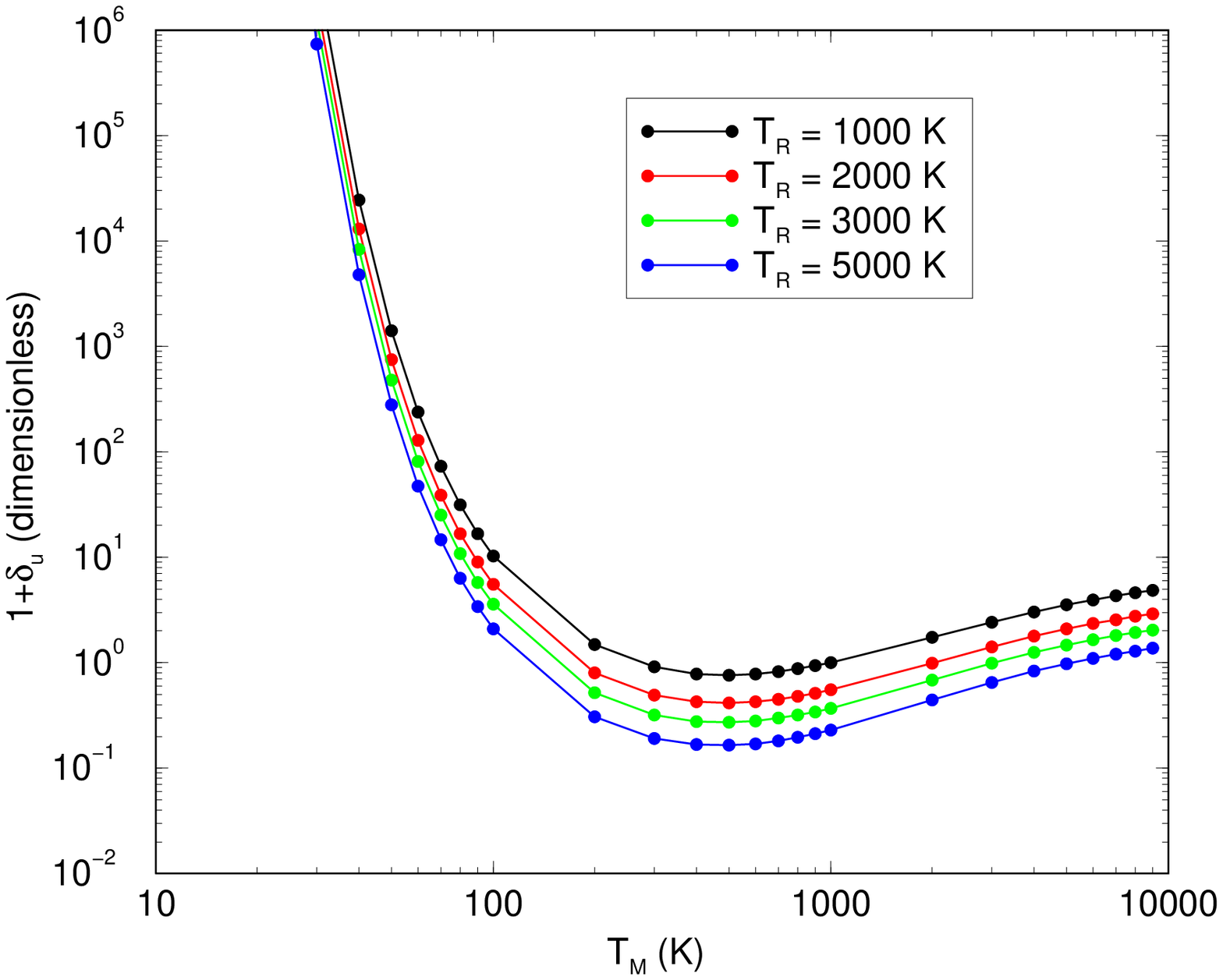}}
\caption{Same as Figure \ref{fig2} except for $j$=29.}
\label{fig3}
\end{figure}

\newpage

\begin{figure}
\centerline{\epsfxsize=4in\epsfysize=2.6in\epsfbox{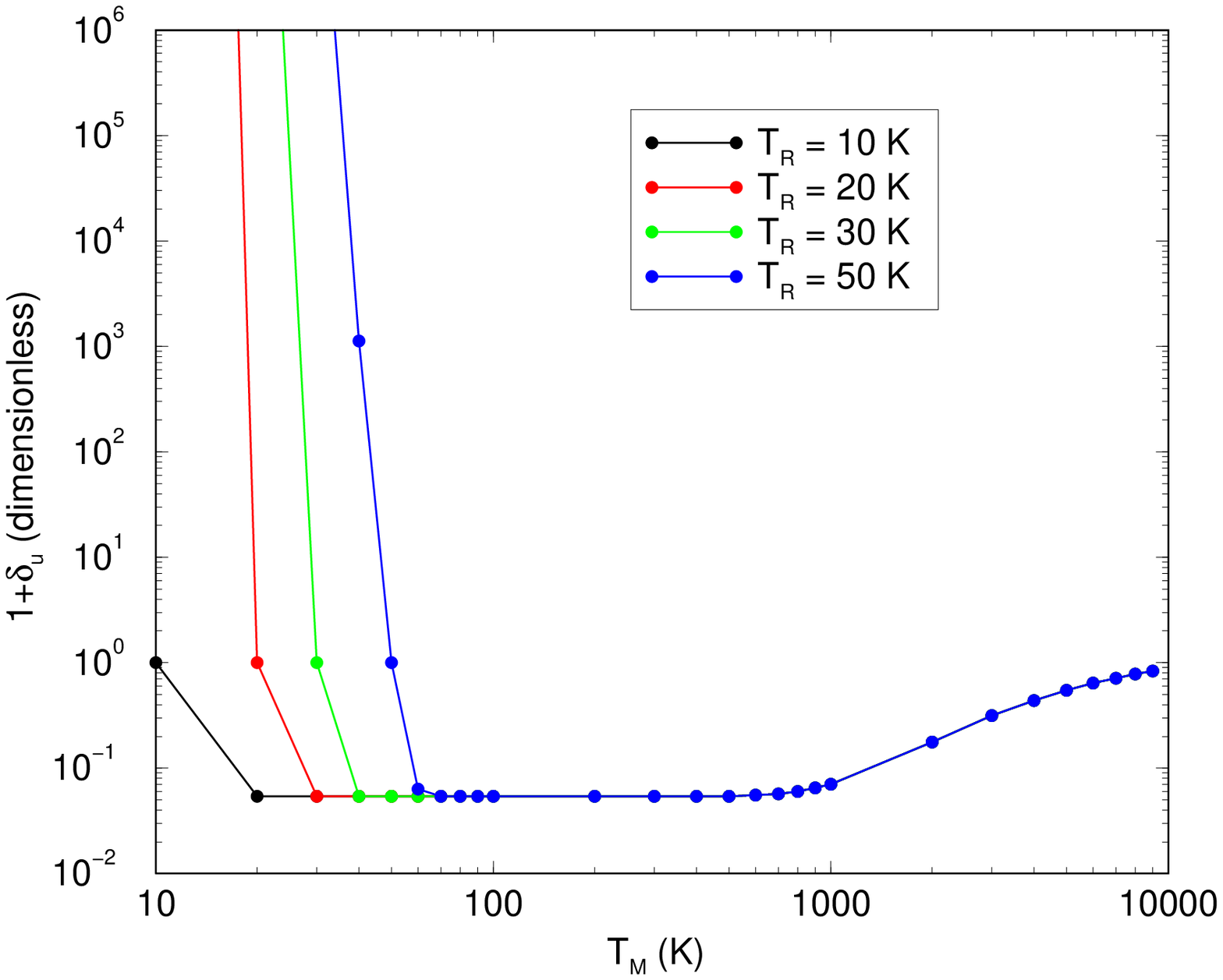}}
\centerline{\epsfxsize=4in\epsfysize=2.6in\epsfbox{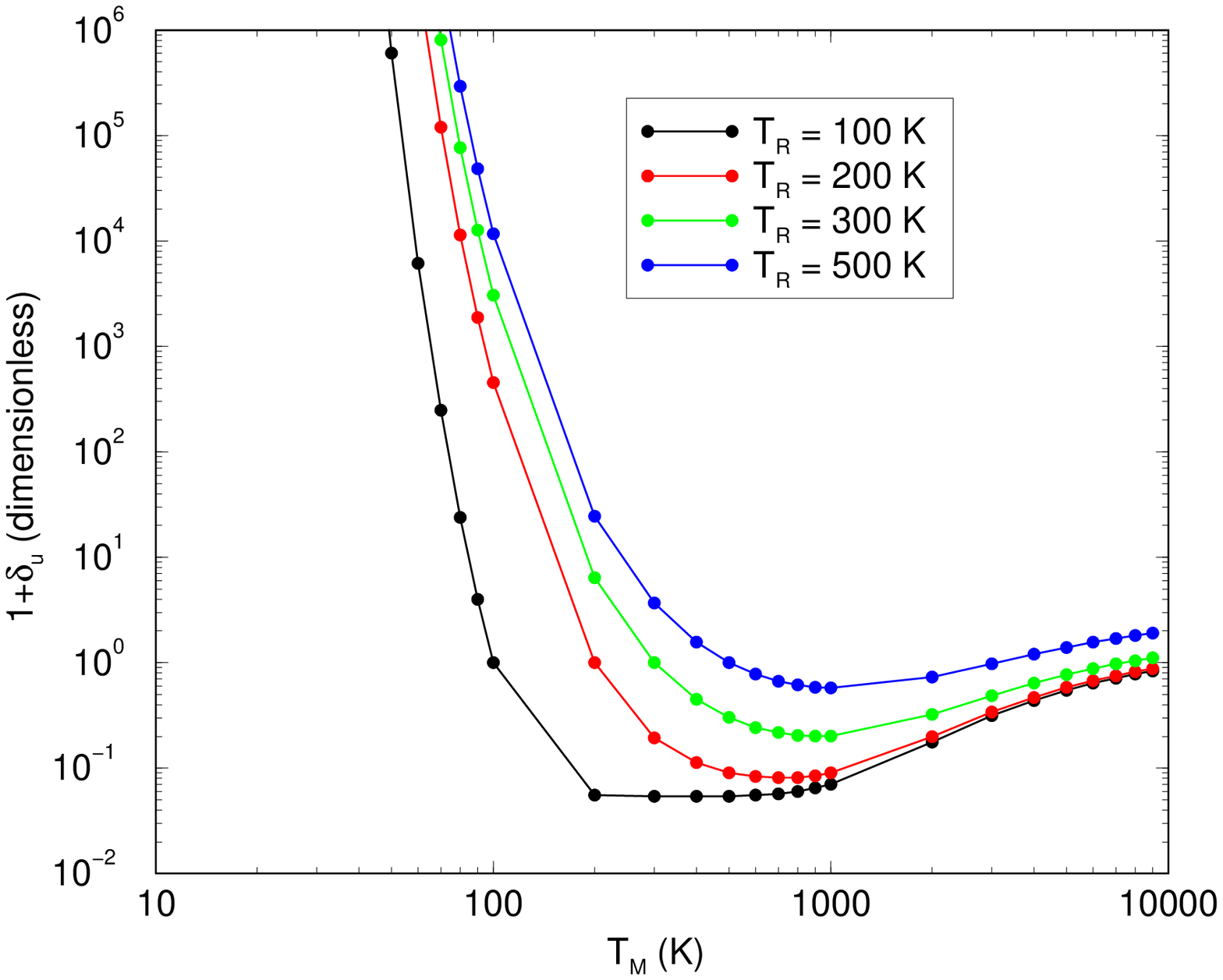}}
\centerline{\epsfxsize=4in\epsfysize=2.6in\epsfbox{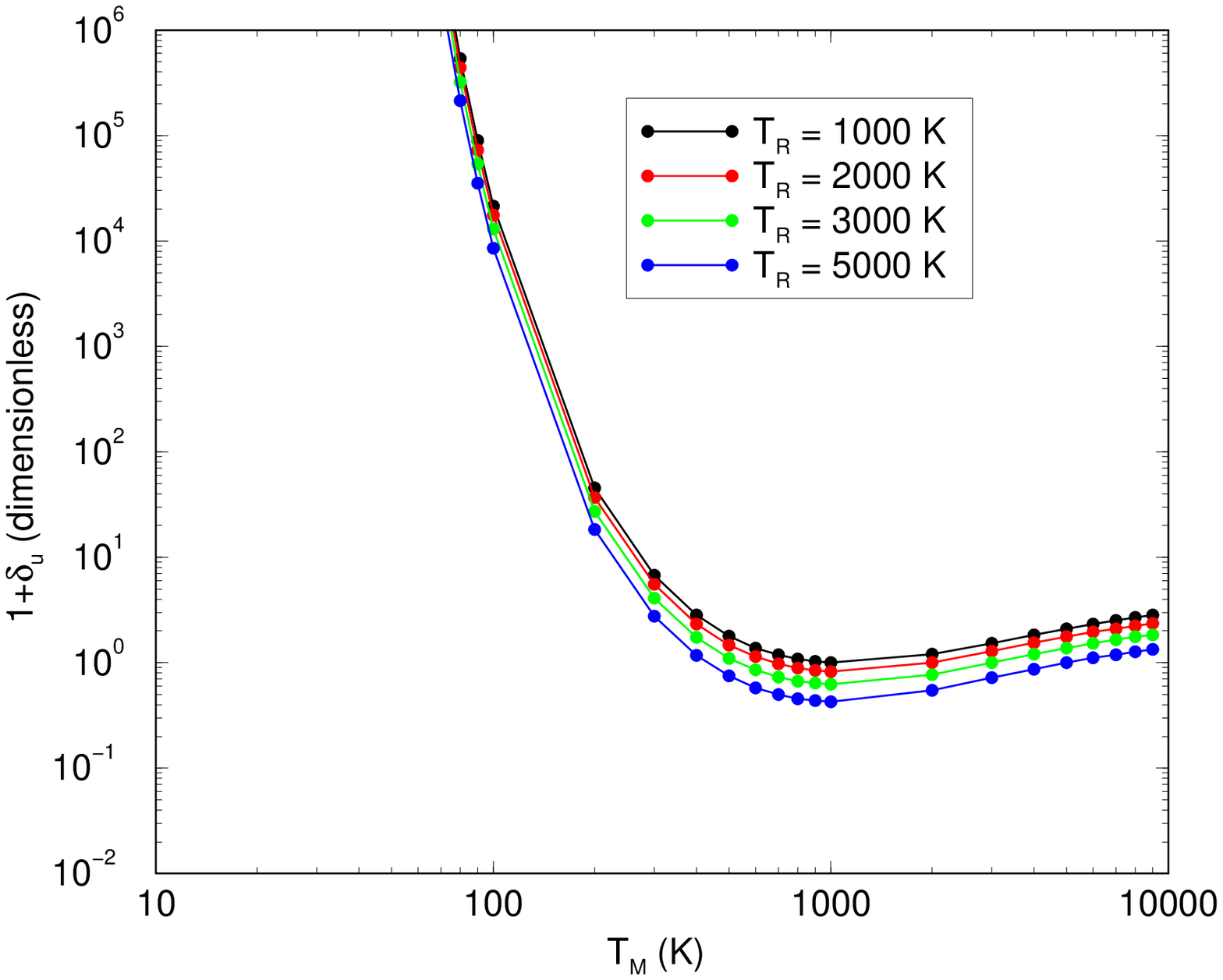}}
\caption{Same as Figure \ref{fig2} except for $j$=31.}
\label{fig4}
\end{figure}

\newpage

\begin{figure}
\centerline{\epsfxsize=4in\epsfysize=2.6in\epsfbox{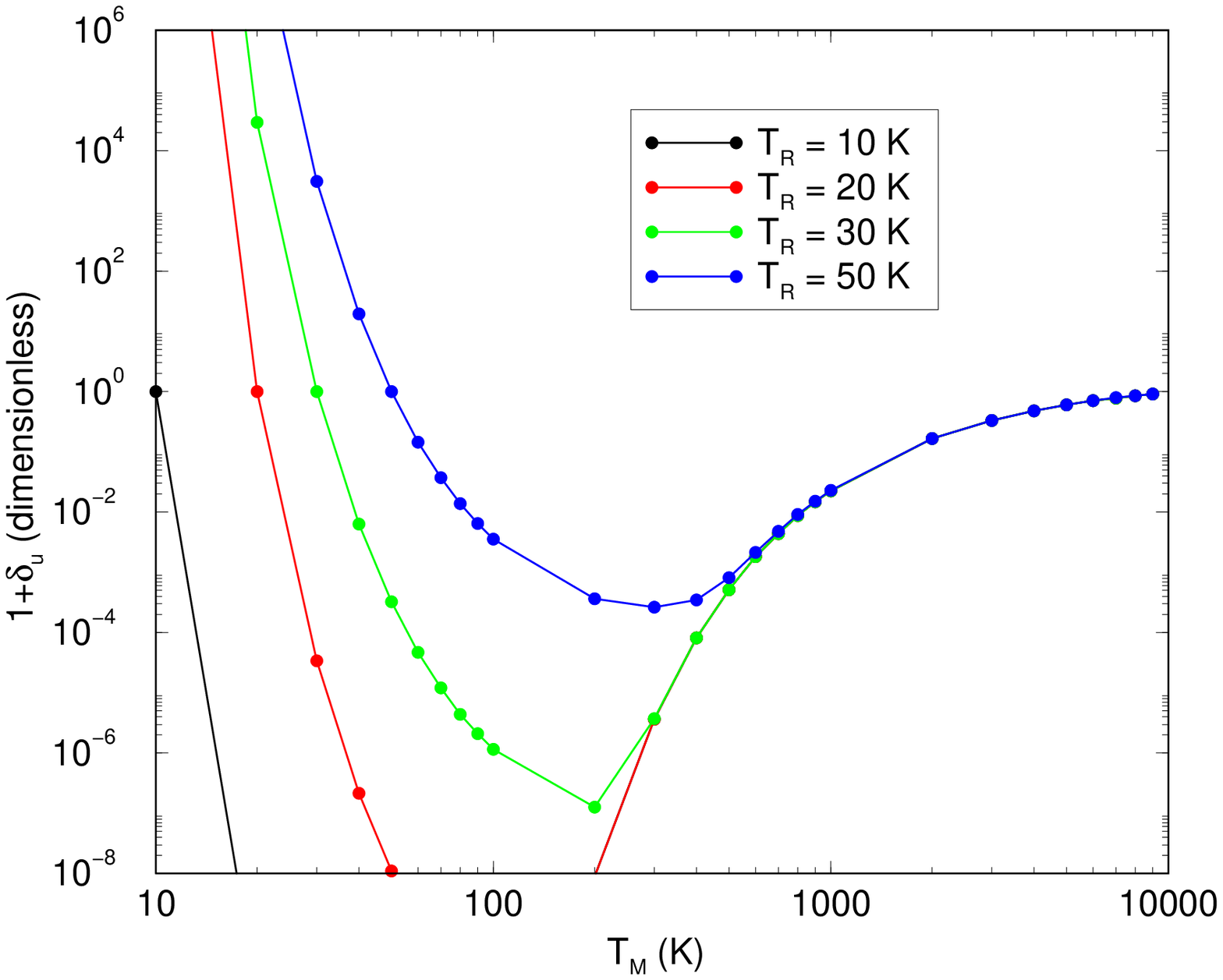}}
\centerline{\epsfxsize=4in\epsfysize=2.6in\epsfbox{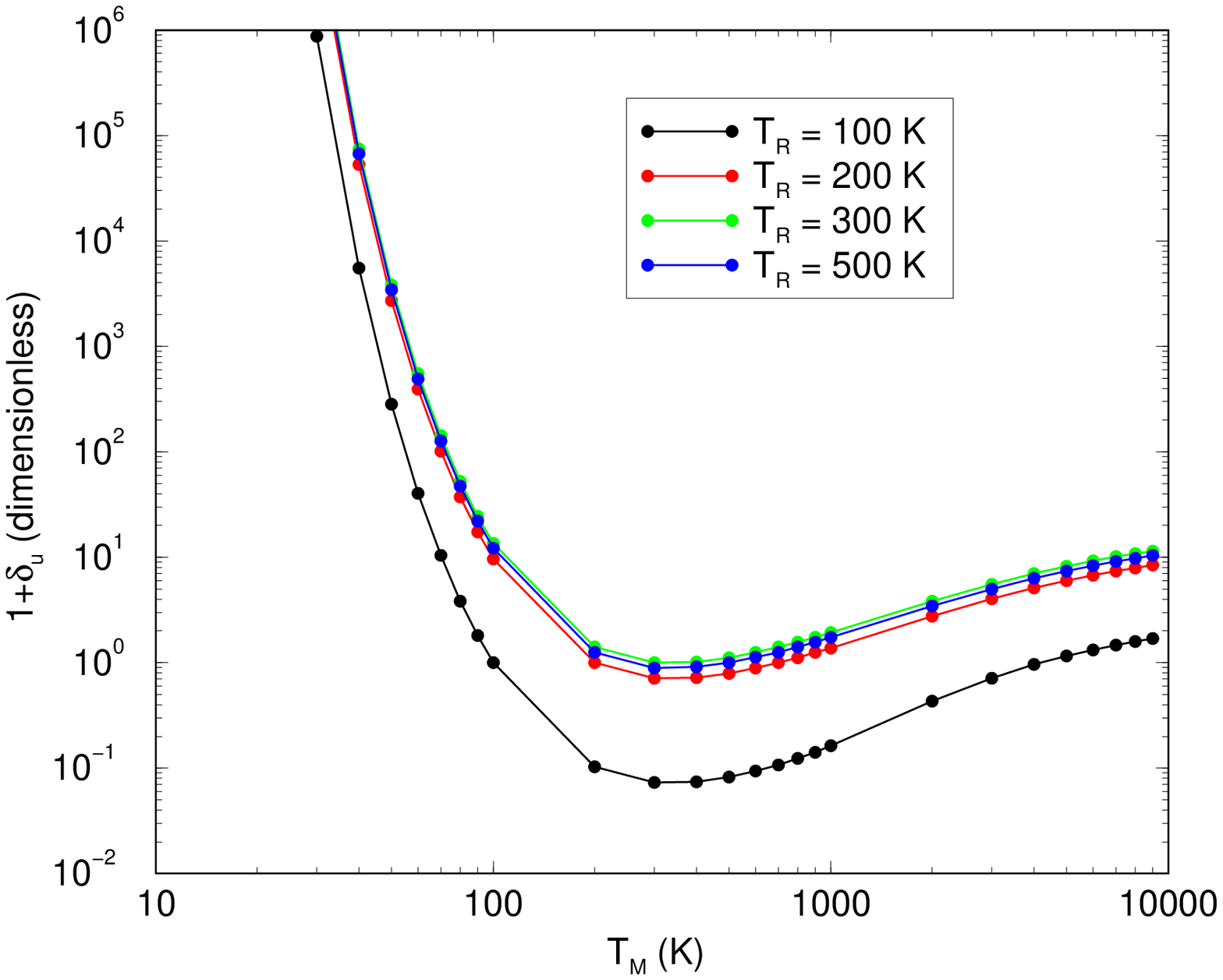}}
\centerline{\epsfxsize=4in\epsfysize=2.6in\epsfbox{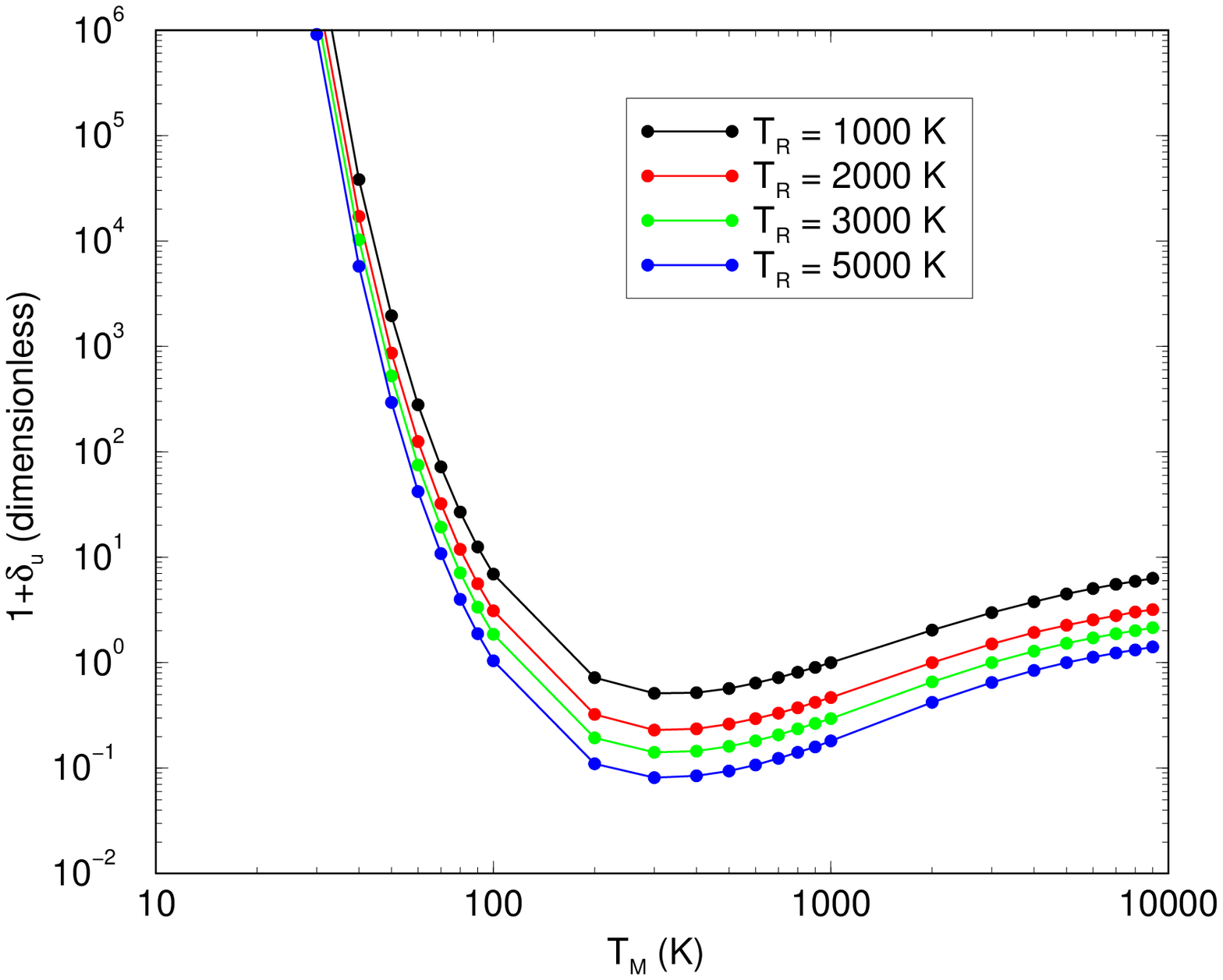}}
\caption{Same as Figure \ref{fig2} except for $j$=32.}
\label{fig5}
\end{figure}

\newpage

\begin{figure}
\centerline{\epsfxsize=4in\epsfysize=2.6in\epsfbox{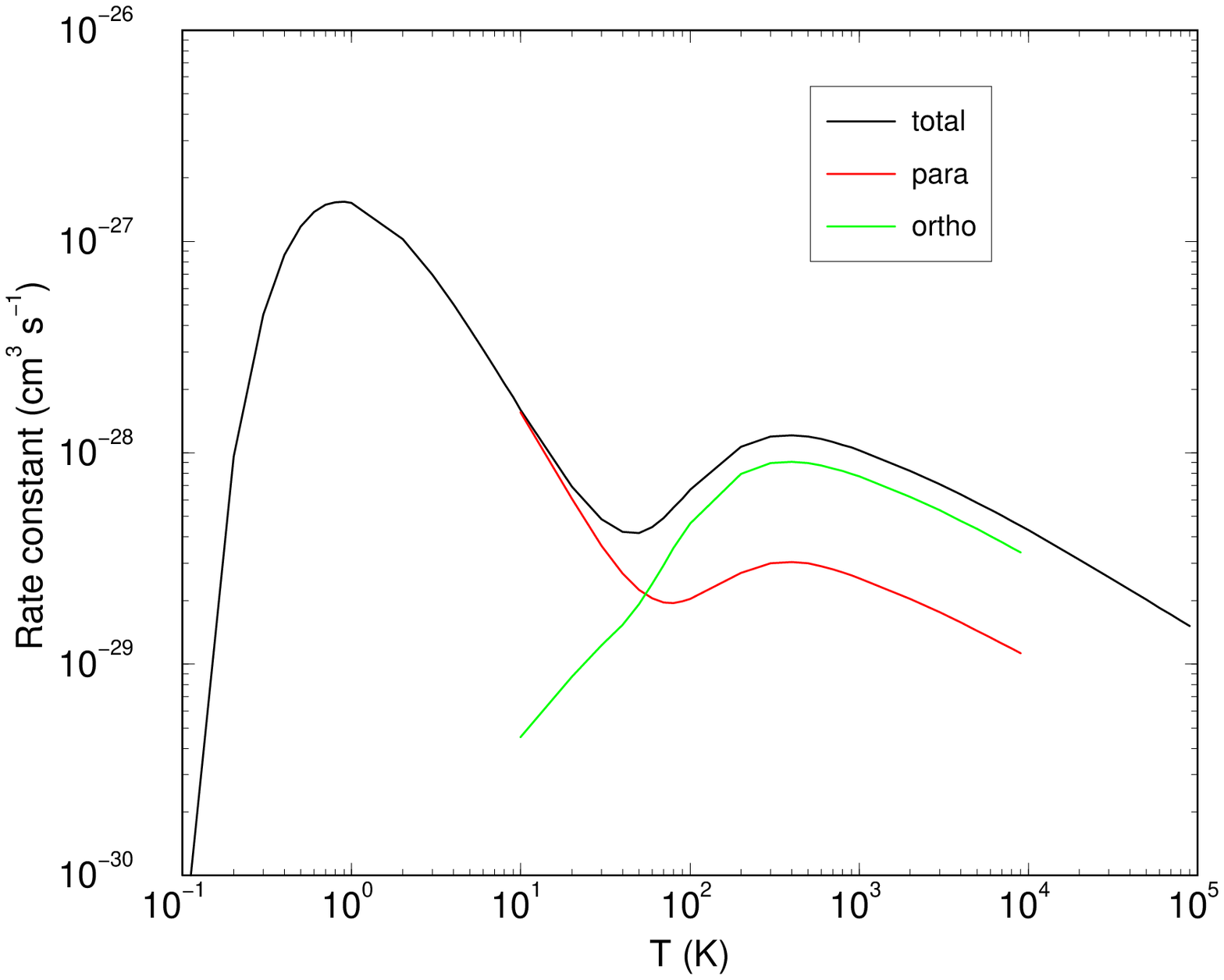}}
\centerline{\epsfxsize=4in\epsfysize=2.6in\epsfbox{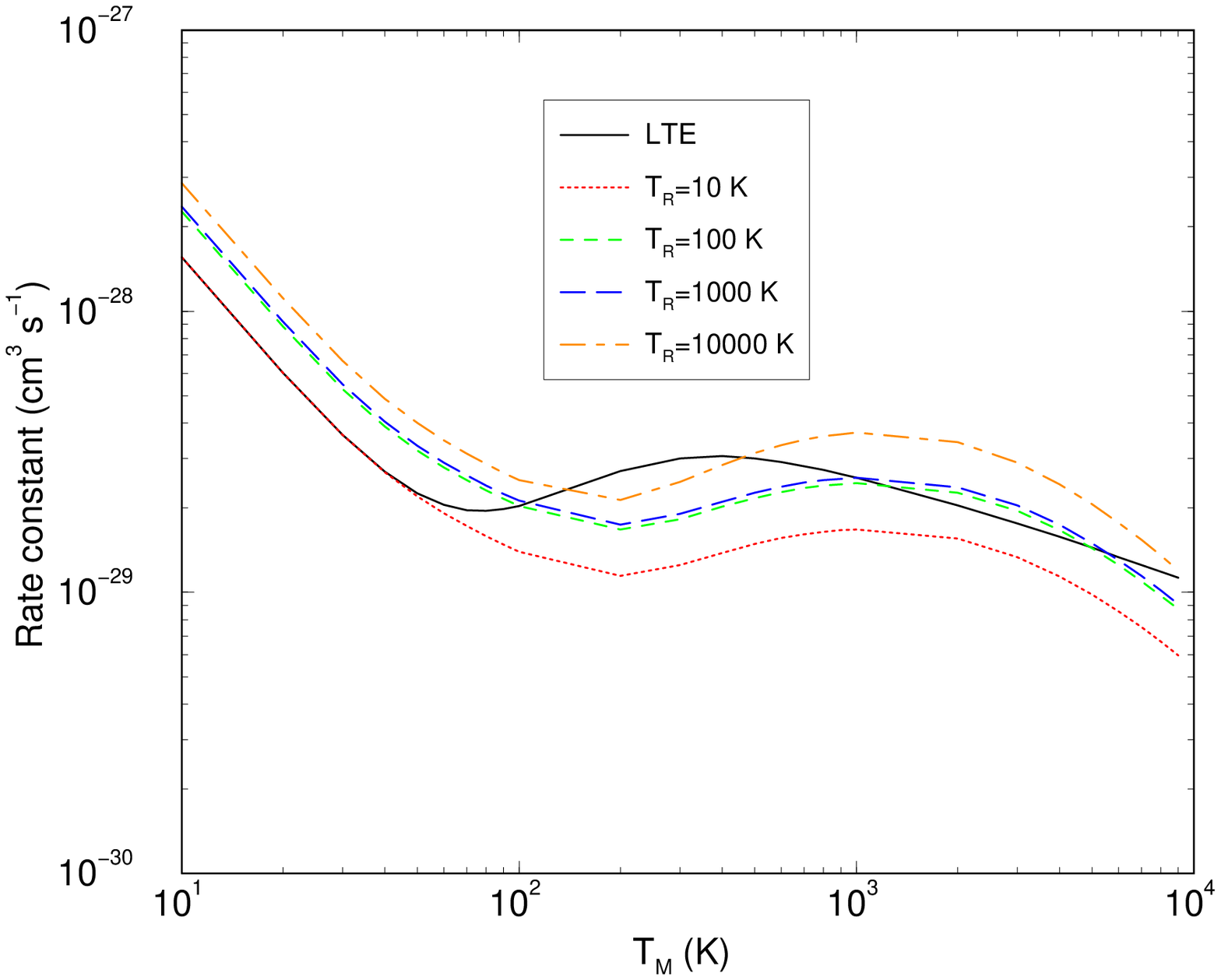}}
\centerline{\epsfxsize=4in\epsfysize=2.6in\epsfbox{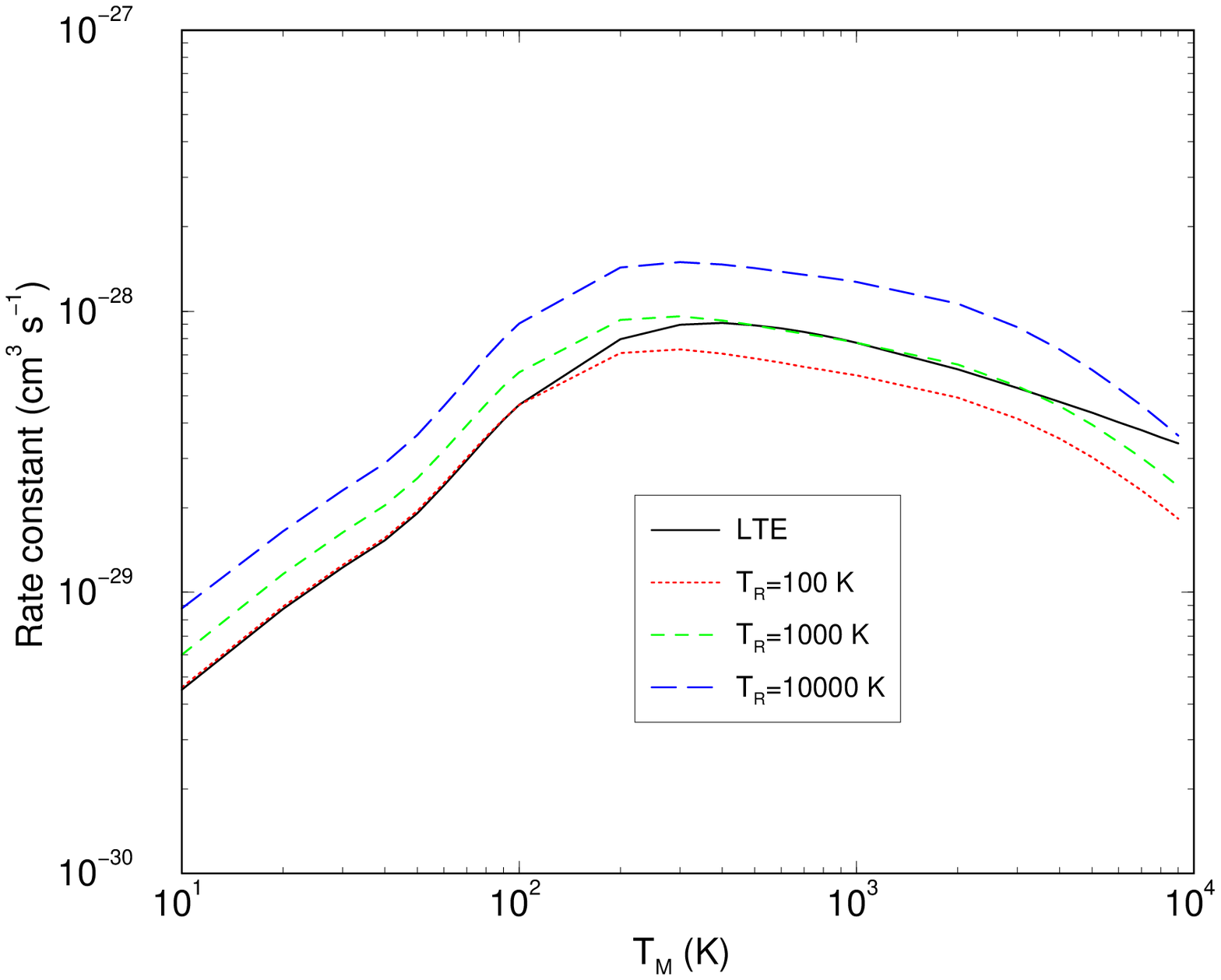}}
\caption{Rate constant for radiative association of H$_2$
at LTE (top panel). Also shown are non-LTE rate constants
as a function of $T_R$ and $T_M$ for forming para-H$_2$ 
(middle panel) and ortho-H$_2$ (bottom panel).
}
\label{fig6}
\end{figure}

\newpage

\begin{figure}
\centerline{\epsfxsize=4in\epsfbox{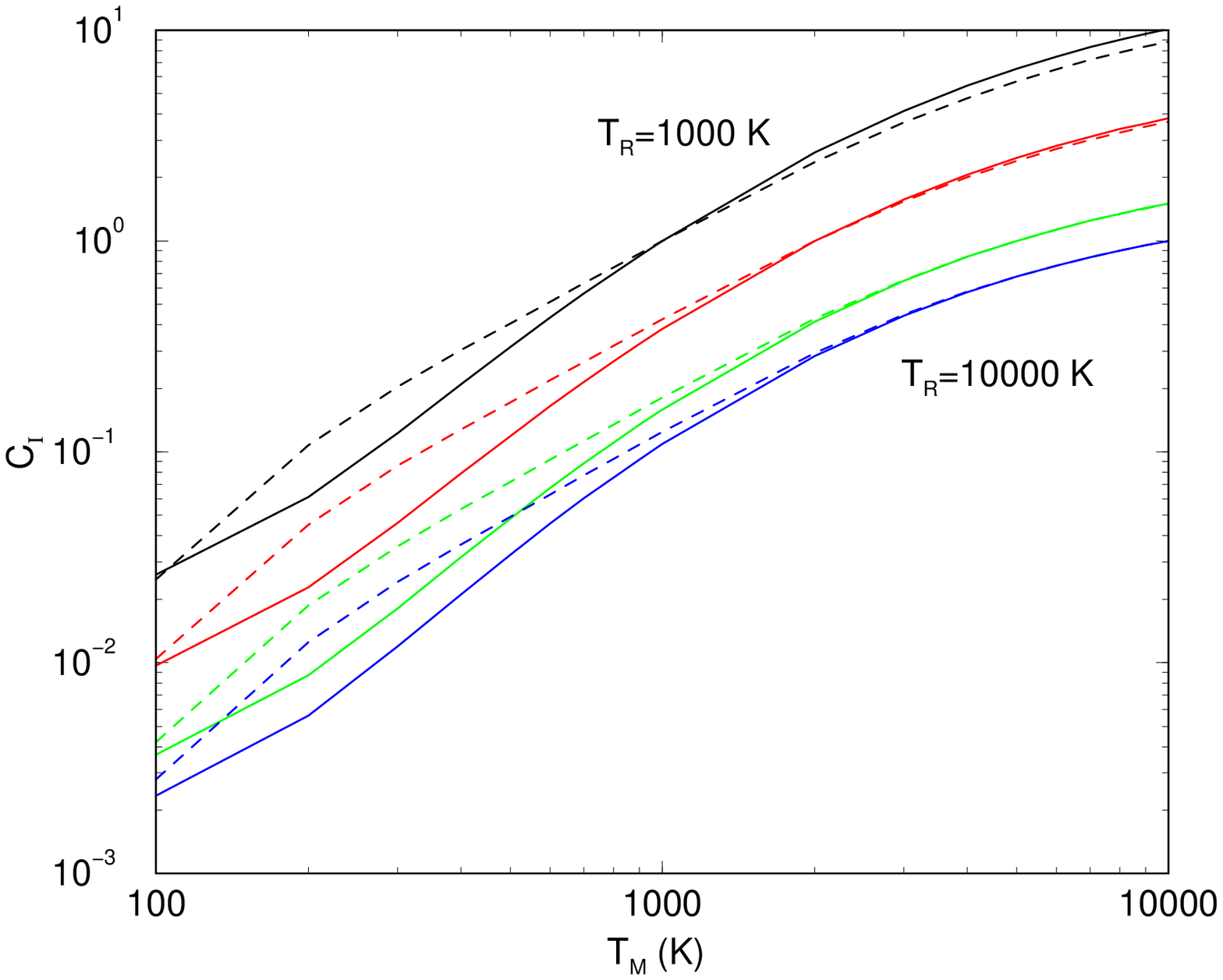}}
\centerline{\epsfxsize=4in\epsfbox{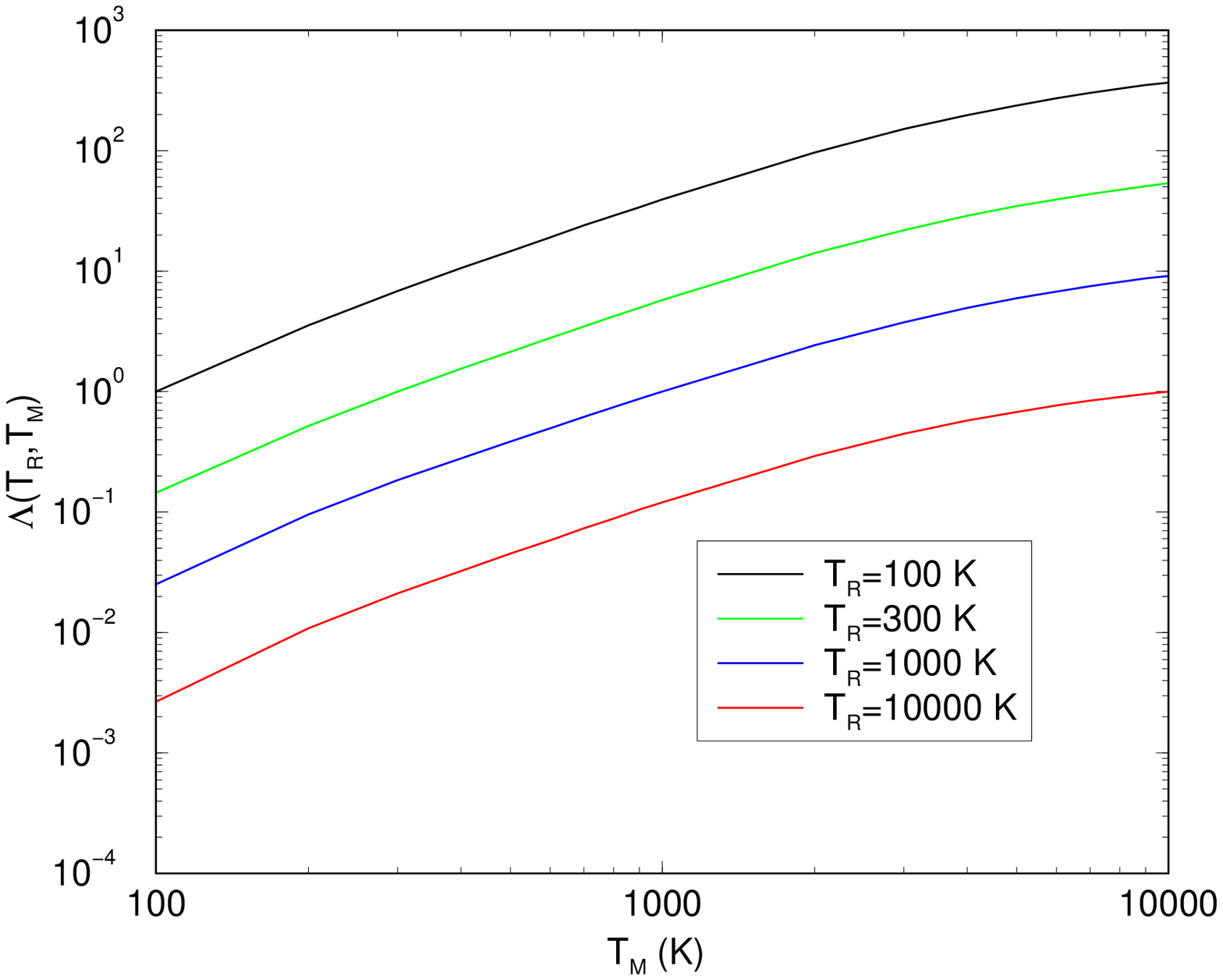}}
\caption{Symmetry constants $C_I$ (top panel)
and scale factor $\Lambda$ (bottom panel).
The symmetry constants are plotted 
for $T_R=1000, 2000, 5000,$ and  10000 K using solid lines 
for $I=0$ and dashed lines for $I=1$. The scale factor is
unity for $T_R=T_M$ but otherwise shows strong variation
due to the assumed absence of inelastic three-body collisions.
}
\label{fig7}
\end{figure}

\newpage

\begin{figure}
\centerline{\epsfxsize=4in\epsfbox{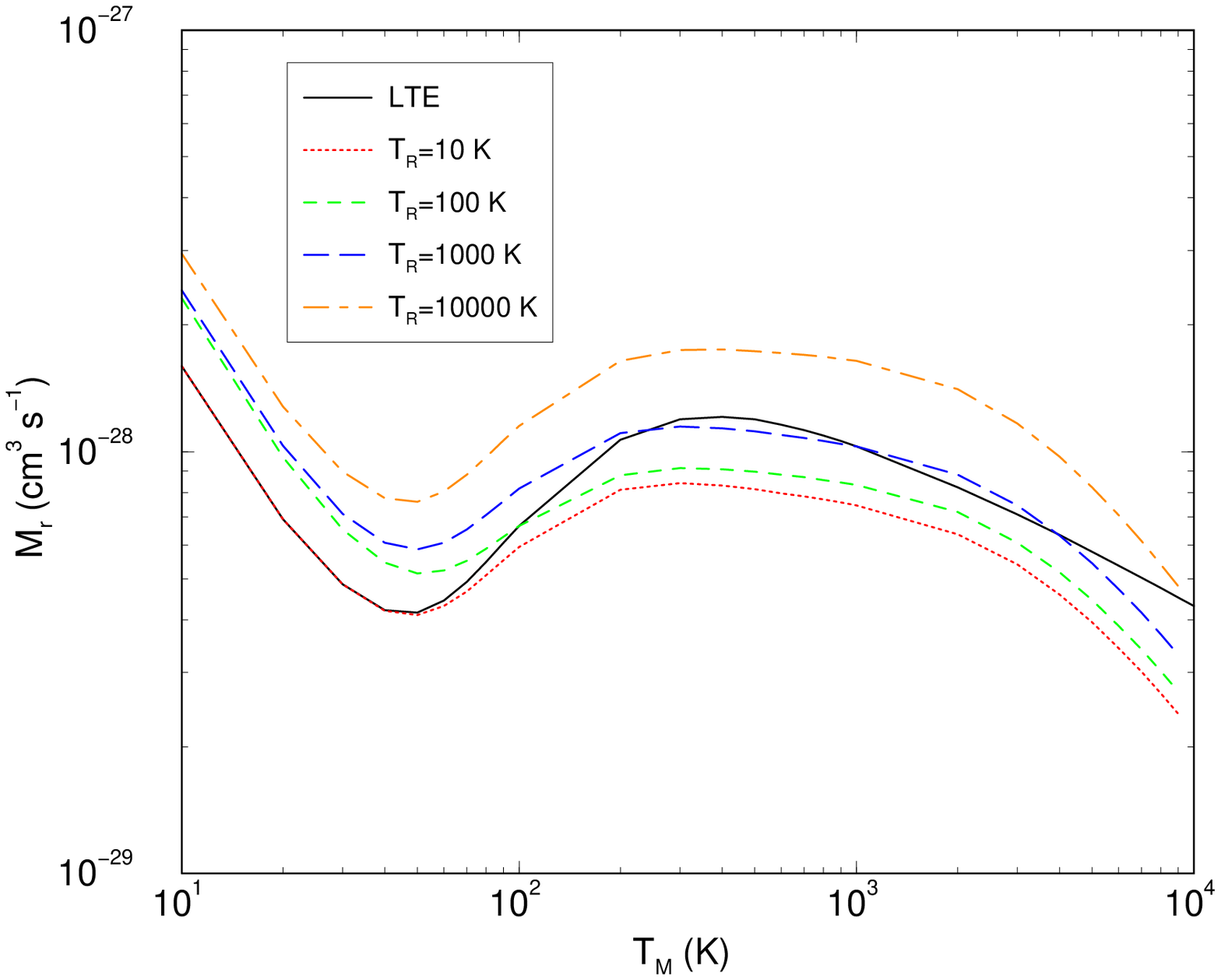}}
\centerline{\epsfxsize=4in\epsfbox{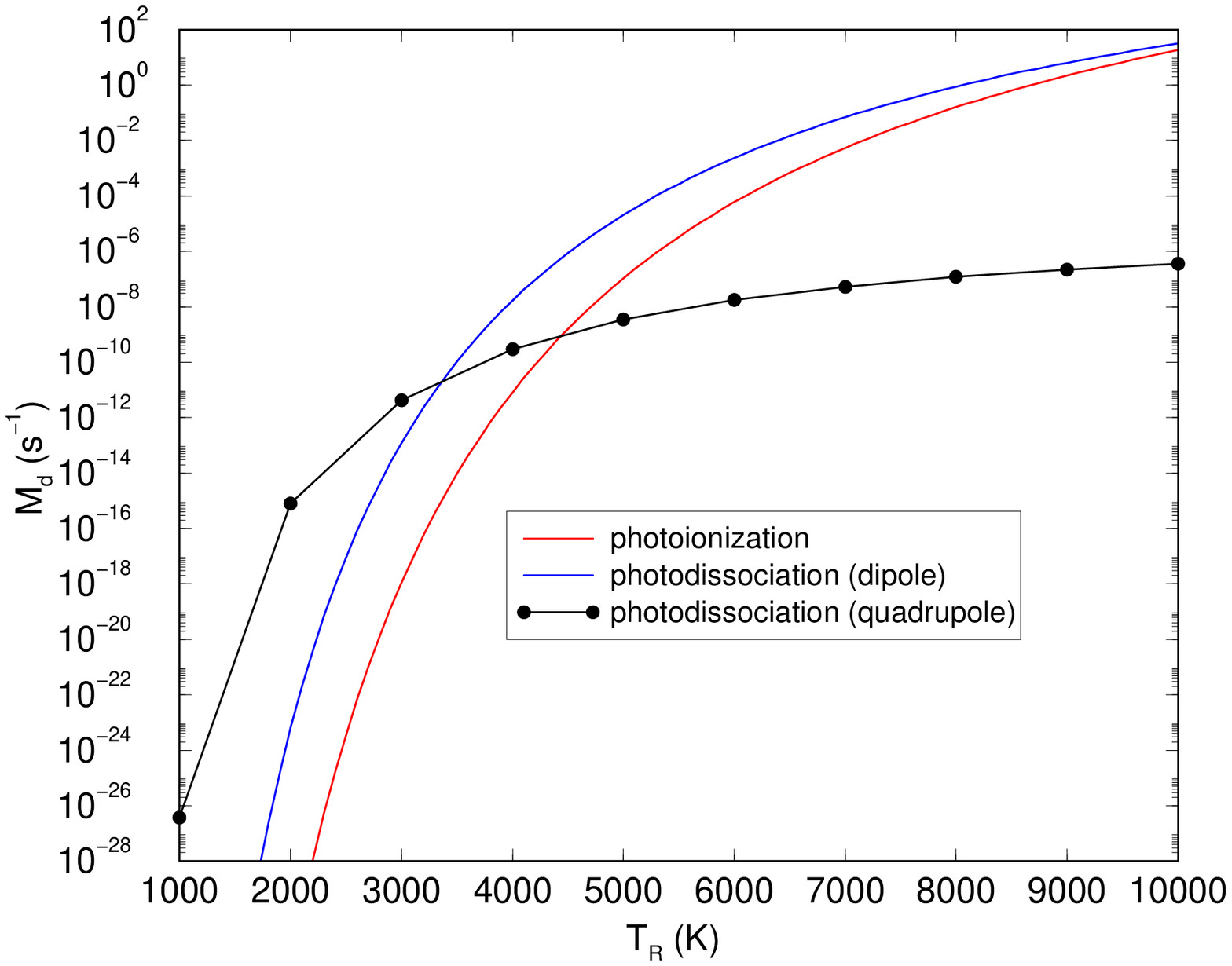}}
\caption{Steady-state rate constants $M_r$ (top panel)
and $M_d$ (bottom panel).
The dissociation rate constant
has a very weak dependence on $T_M$ and
is plotted as a function of $T_R$. Also shown on the
bottom panel are rate constants for photoionization
and indirect photodissociation (dipole) obtained from
\cite{coppola2011}.  }
\label{fig8}
\end{figure}

\newpage

\begin{figure}
\centerline{\epsfxsize=4in\epsfbox{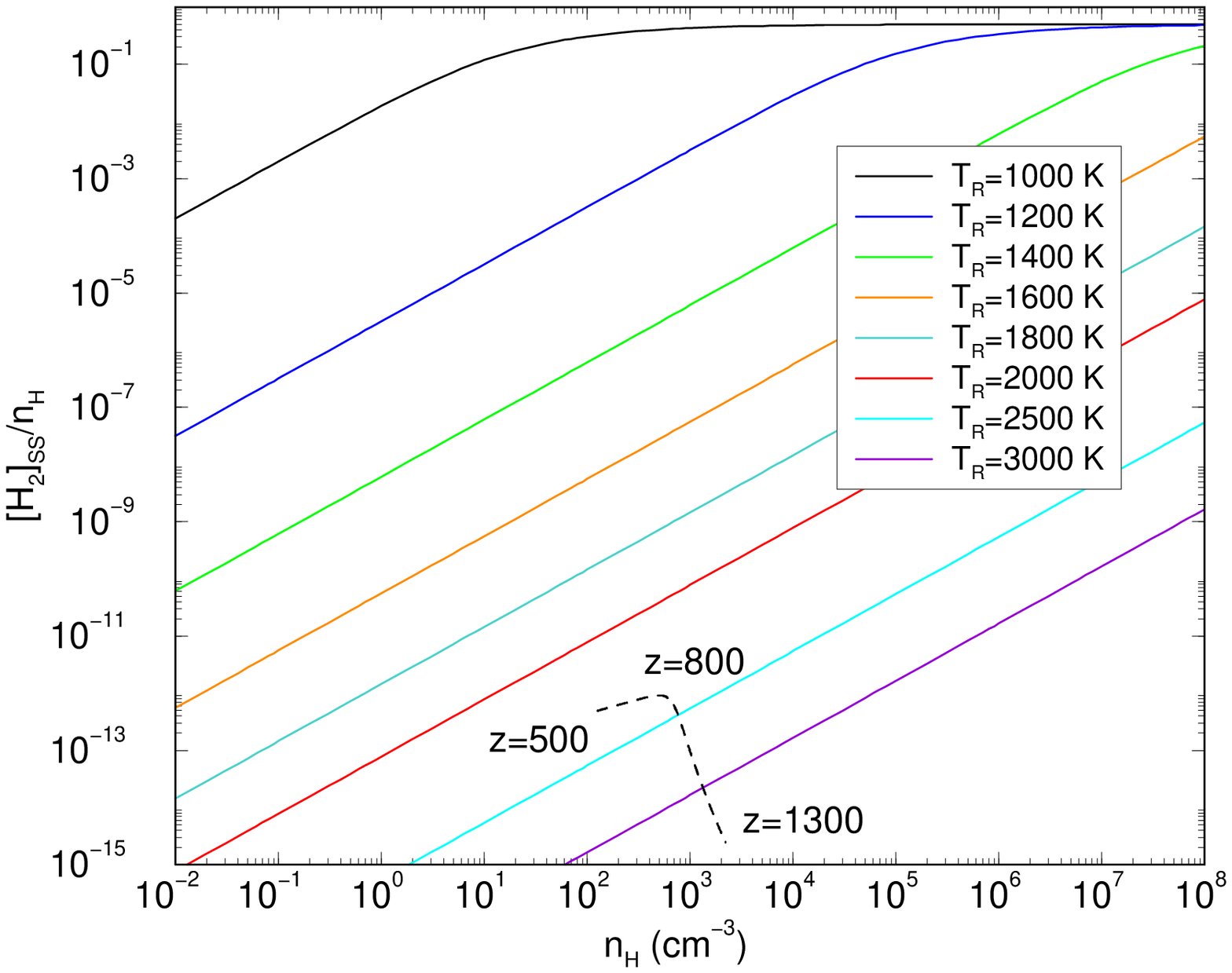}}
\centerline{\epsfxsize=4in\epsfbox{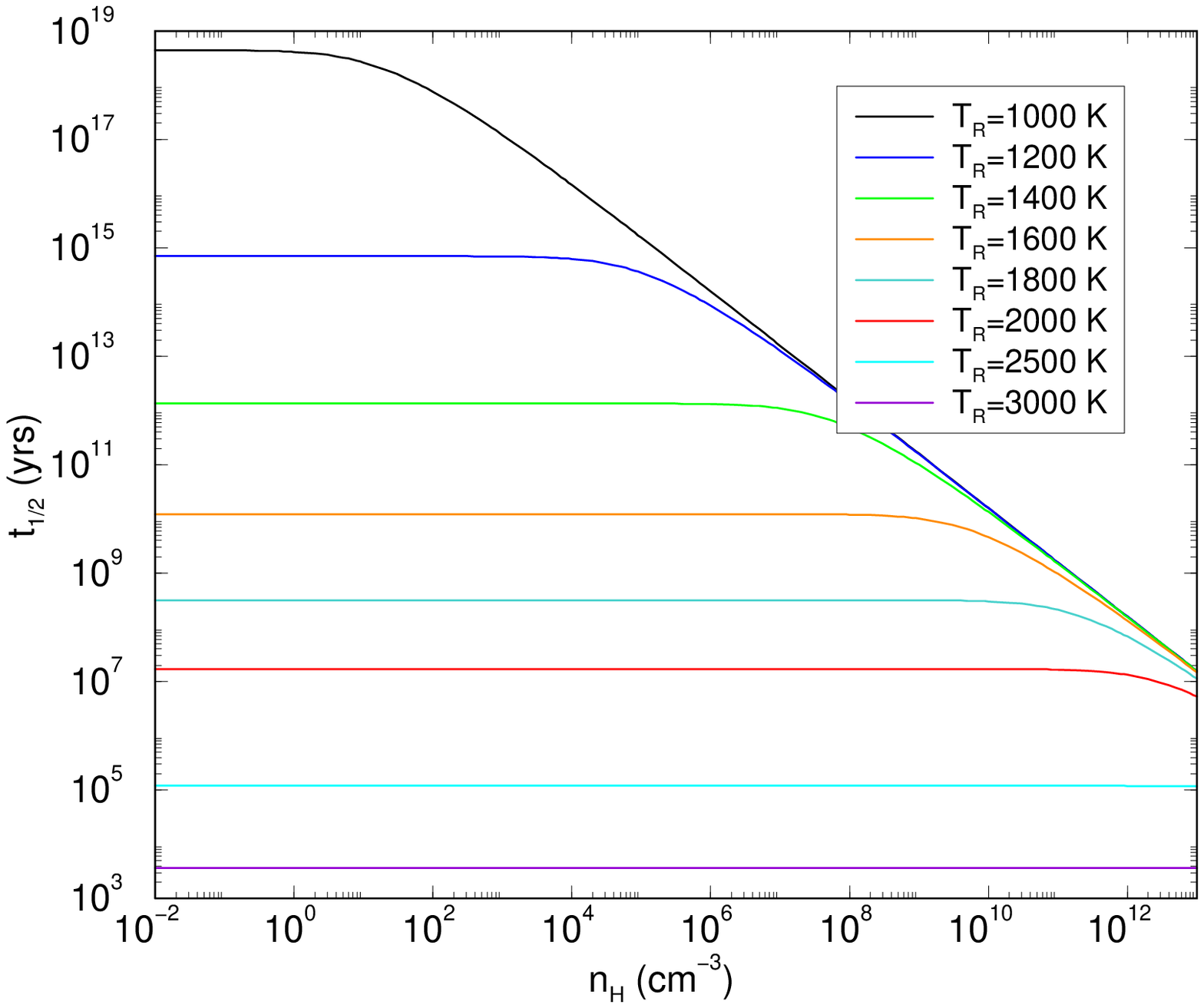}}
\caption{Steady-state fractional population of H$_2$ (top panel)
and the time required to reach one-half the steady-state value
(bottom panel) for a system which considers only quadrupole 
association and dissociation of hydrogen.
When $T_R<2500$ K, the system is not able to reach steady-state
on a realistic time scale. In this case, the time-dependent
solution (\ref{H2}) is shown as a dashed curve on the top
panel for $500<z<1300$.  }
\label{fig9}
\end{figure}

\end{document}